\documentclass{article}
\usepackage{epsfig, graphics}
\usepackage{lscape}
\usepackage{multicol}
\usepackage{amssymb,amstext}

\def\squareforqed{\hbox{\rlap{$\sqcap$}$\sqcup$}}
\def\qed{\ifmmode\squareforqed\else{\unskip\nobreak\hfil
\penalty50\hskip1em\null\nobreak\hfil\squareforqed
\parfillskip=0pt\finalhyphendemerits=0\endgraf}\fi}

\newcommand{\ab}{\allowbreak }

\newcommand{\lland}{\mathrel{\land}}

\newcommand{\R}{{\rm {\bf R}}}

\newcommand{\faff}{${\cal F}_{\rm Aff}$}
\newcommand{\fsc}{${\cal F}_{\rm Sc}$}
\newcommand{\ftrans}{${\cal F}_{\rm Trans}$}
\newcommand{\fid}{${\cal F}_{\rm id}$}

\newcommand{\faffrat}{${\cal F}_{\rm Aff}^{\rm Rat}$}
\newcommand{\fscrat}{${\cal F}_{\rm Sc}^{\rm Rat}$}
\newcommand{\ftransrat}{${\cal F}_{\rm Trans}^{\rm Rat}$}
\newcommand{\faffl}{${\cal F}_{\rm Aff}^{\rm L}$}
\newcommand{\fscl}{${\cal F}_{\rm Sc}^{\rm L}$}
\newcommand{\ftransl}{${\cal F}_{\rm Trans}^{\rm L}$}

\newcommand{\fafflin}{${\cal F}_{\rm Aff}^{\rm Lin}$}
\newcommand{\fsclin}{${\cal F}_{\rm Sc}^{\rm Lin}$}
\newcommand{\ftranslin}{${\cal F}_{\rm Trans}^{\rm Lin}$}
\newcommand{\faffpoly}{${\cal F}_{\rm Aff}^{\rm Poly}$}
\newcommand{\fscpoly}{${\cal F}_{\rm Sc}^{\rm Poly}$}
\newcommand{\ftranspoly}{${\cal F}_{\rm Trans}^{\rm Poly}$}

\newcommand{\spoly}{${\cal S}_{\rm Poly}$}
\newcommand{\srect}{${\cal S}_{\rm Rect}$}
\newcommand{\str}{${\cal S}_{\rm Tr}$}
\newcommand{\strax}{${\cal S}_{\rm TrAx}$}

\newcommand{\sspoly}{{\cal S}_{\rm Poly}}
\newcommand{\ssrect}{{\cal S}_{\rm Rect}}
\newcommand{\sstr}{{\cal S}_{\rm Tr}}
\newcommand{\sstrax}{{\cal S}_{\rm TrAx}}

\newtheorem{definition}{Definition}[section]
\newtheorem{theorem}{Theorem}[section]

\newtheorem{lemma}{Lemma}[section]

\newtheorem{property}{Property}[section]

\begin{document}
\title{Classes of Spatiotemporal Objects \\ and Their Closure Properties}

\author{Jan Chomicki \\ {\small University at Buffalo, SUNY} \\ {\small
Department of Computer}\\{\small Science and Engineering} \\
{\small 201 Bell Hall Box 602000}\\{\small Buffalo, NY 14260-2000,
US}\\ {\small {\tt chomicki@cse.buffalo.edu}} \and Sofie
Haesevoets
\\ {\small University of Limburg}\\ {\small Department of \ Mathematics,}\\ {\small Physics and Computer
Science}\\{\small Universitaire Campus}\\ {\small B-3590
Diepenbeek, Belgium}\\ {\small {\tt sofie.haesevoets@luc.ac.be}}
\and Bart
Kuijpers%\footnote{Contact author.}
\\ {\small University of Limburg}\\ {\small Department of \ Mathematics,}\\ {\small Physics and Computer
Science}\\{\small Universitaire Campus}\\ {\small B-3590
Diepenbeek, Belgium}\\ {\small {\tt bart.kuijpers@luc.ac.be}} \and
Peter Revesz\\ {\small University of Nebraska-Lincoln}\\ {\small
Department of Computer }\\ {\small Science and Engineering}\\
{\small Ferguson 214}\\ {\small Lincoln, NE 68588-0115, US}\\
{\small {\tt revesz@cse.unl.edu}}}

\date{}

\maketitle

\begin{abstract}
We present a data model for spatio-temporal databases. In this
model spatio-temporal data is represented  as a finite union of
objects described by means of a spatial reference object, a
temporal object and a geometric transformation function that
determines the change or movement of the reference object in time.

We define a number of practically relevant classes of
spatio-temporal objects, and give complete results concerning
closure under Boolean set operators for these classes. Since only
few classes are closed under all set operators, we suggest an
extension of the model, which leads to better closure properties,
and therefore increased practical applicability. We also discuss a
normal form for this extended data model.
\end{abstract}

\thispagestyle{empty}

\smallskip

\section{Introduction}\label{intro}  Many natural or man-made phenomena have both
a spatial and a temporal extent. Consider for example a forest
fire, a meteorological event (e.g., the movement of clouds
pressure areas), property histories in a city or the flight of an
air plane. To store information about such phenomena in a
database, appropriate data modeling constructs are needed.

In this paper, we introduce and discuss a general framework for
specifying spatio-temporal data. Hereto, the new concept of
spatio-temporal object is introduced. We represent a
spatio-temporal object as a finite number of objects represented
by means of a spatial reference object, a temporal object (i.e., a
time interval) and a time-dependent geometric transformation that
determines how this spatial object moves or changes through space
during the considered time interval. Although this model is suited
for data in arbitrary dimensions, we focus on two-dimensional
reference objects that move or change during time.

In this framework, a number of classes of practically relevant
spatio-temporal objects arise naturally. These classes are indexed
by the type of spatial reference object and the type of
transformation functions that are allowed. On the level of
reference objects, we consider polygons, triangles, triangles with
two sides parallel to the coordinate axes of the two-dimensional
plane and rectangles with all sides parallel to the coordinate
axes. We consider time-dependent affinities, scalings and
translations for what concerns transformation functions. These
functions can be expressed by rational, polynomial, respectively
linear functions.

We investigate these classes with respect to closure under Boolean
set operations, namely union, intersection and set-difference.

By definition, these classes are closed under union (a
spatio-temporal object is described as the union of atomic
objects). We call a class closed under intersection (respectively
set-difference) if any finite intersection (respectively
set-difference) of objects from a class can again  be described by
an object from that class (i.e., as a union of atomic objects).
The classes that we consider are not necessarily closed under
intersection and set-difference.

We provide an in-depth and exhaustive study of their closure with
respect to all set-theoretic operations, and we conclude  that our
model for representing spatio-temporal data gives very poor
closure results for the classes of objects we considered important
for spatio-temporal practice. The only classes that seems to be
useful in this respect have polygons as spatial reference objects
and use rational affinities to move or change these objects in
time.

A conclusion is that we have to enrich the data model by allowing
set-theoretic operations other than union in the construction of
geometric objects from atomic geometric objects. As soon as we
also allow spatio-temporal objects to be constructed from atomic
ones by means of union and intersection (or union and
set-difference) then the model becomes closed for {\em all}
Boolean set operations. Indeed, as an important result, we show
that our classes, that are drawn from practice, have the nice
property that they are closed under intersection if and only if
they are closed under set-difference.

To appreciate the need for applying set-theoretic operators to
spatiotemporal objects, consider the following scenario. Let two
spatial objects represent the extents of the safe areas around two
different ships. Taking into account the movement of ships, the
extents of the safe areas over a period of time can be represented
as two spatiotemporal objects. To avoid collisions, one needs to
be able to determine the intersection of those objects.

The substantial literature on spatial and temporal databases does
not provide much guidance in dealing with spatiotemporal
phenomena. Spatial databases~\cite{Wor95} deal with spatial
objects (e.g., rectangles or polygons) and temporal
databases~\cite{TDB93} with temporal ones (e.g., time intervals).
Their combination can handle {\em discrete} change~\cite{Wor94}
but not {\em continuous} change, which is required by applications
dealing with phenomena like movement, natural disasters, or the
growth of urban areas. In the latter applications, the temporal
and spatial aspects cannot be conveniently separated.

Spatiotemporal data models and query languages are a topic of
growing interest. The need to model both discrete and continuous
change has been identified. The issue of closure under Boolean set
operations has also received in this context a considerable
attention. This is not surprising, since, for example, closure
under intersection is essential for {\em spatiotemporal join}.

The paper~\cite{Wor94} presents one of the first such models.
However, it is only capable of modeling discrete change.

In \cite{ErGuScVa99} the authors define in an abstract way {\em
moving points and regions}. Apart from moving points, no other
classes of concrete, database-representable spatiotemporal objects
are defined. In that approach continuous movement (but not growth
or shrinking) can be modeled using linear interpolation functions.
In the subsequent paper \cite{FoGuNaSc00}, the authors discuss a
concrete, polyhedral representation of {\em moving}, {\em growing}
and {\em shrinking regions}, which is applicable only to
significantly restricted classes of spatiotemporal objects.  This
guarantees closure but eliminates the possibility of representing
scaling and more general transformations. The results of the
present paper shed some light on when similar concrete
representations exist and when they do not.

In \cite{GrRiSeGIS98} the authors propose a formal spatiotemporal
data model based on {\em constraints} in which, like
in~\cite{Wor94}, only discrete change can be modeled.  An
SQL-based query language is also presented.

We have proposed elsewhere \cite{ChReGEO99} a spatiotemporal data
model based on {\em parametric polygons}: polygons whose vertices
are defined using linear functions of time. This model is also
capable of modeling continuous change but is not closed under
intersection.  A variation of this model restricted to rectangles
but extended with periodic functions is given in~\cite{CKR00}.
The latter model is closed under set theoretic operators, enabling
the definition of an extended relational algebra query language,
for which query evaluation can be done in PTIME in the size of the
input spatiotemporal database.  The closure properties for
\cite{ChReGEO99} and \cite{CKR00} seem analoguous to the closure
properties of the framework presented in this paper, respectively,
but the relationships among these frameworks needs to be further
explored.

Both discrete and continuous change can be represented using {\em
constraint databases} \cite{KaKuRe95}.  Compared to the latter
technology, our approach seems more constructive and amenable to
implementation using standard database techniques.  On the other
hand, constraint databases do not suffer from the lack of closure
under intersection.  To some degree, it is due to the fact that
the intersection of two generalized tuples in constraint databases
need not immediately computed but rather the tuples may be only
conjoined together.  In most implementations of constraint
databases~\cite{BrSeChEx97,GrRiSeSIGMOD98,ReLi97} the ``real''
computation of the intersection occurs during projection or the
presentation of the query result to the user. It is unclear
whether such a strategy offers any computational advantages over
the approach in which the intersections are computed immediately.
In fact, recent work on spatial constraint databases
\cite{KuRaShSu98} proposes extensions to relational algebra that
require immediate computations of spatial object intersections.
Also, our approach is potentially more general than constraint
databases.  For example, by moving beyond rational functions (but
keeping the same basic framework) we can represent rotations with
a fixed center.  Finally, in our model it is easy to obtain any
snapshot of a spatiotemporal object, making tasks like animation
straightforward. It is not so in constraint databases where
geometric representations of snapshots have to be explicitly
constructed from constraints~\cite{clr99}.

This paper is organized as follows: In Section 2, we give
definitions and describe the relevant classes of spatio-temporal
objects. The closure results for these classes with respect to
Boolean set operations are given in Section 3. We propose the
extended model in Section 4 and describe a normal form for objects
in this extended model. Section 5 gives comments and concludes the
paper.

\section{Definitions and preliminaries}\label{defs}

In this section, we define the notion of {\em spatio-temporal
object}. In our approach, a spatio-temporal object consists of a
spatial reference object, a time interval during which the
spatio-temporal object exists and a continuous transformation that
defines how the spatial reference object moves and changes during
the interval of time.

\subsection{Spatio-temporal and geometric
objects} Let $\R$ be the set of real numbers and $\R^2$ be the
$2$-dimensional real plane.

\begin{definition}\label{defobj}\rm
A {\em spatial object\/} is a subset of $\R^2$. A {\em temporal
object\/} is a subset of $\R$ (we assume a single temporal
dimension). A {\em spatio-temporal object\/} is a subset of
$\R^{2}\times \R$.\qed
\end{definition}

These definitions are very general and disregard the fact that
objects should be finitely representable in the computer's memory.
In this paper, we will study more restricted classes of spatial
and spatio-temporal objects that are important from a practical
point of view and have simple and efficient representations. Such
classes have been identified in the course of spatial and
spatio-temporal database research.

Here, we propose a geometric approach: a spatio-temporal object is
defined as a spatial reference object together with a continuous
transformation that defines how the object moves or changes during
some time interval.

\begin{definition}\rm
An {\em atomic geometric object $\cal O$\/} is a triple $(S,I,f)$,
where
\begin{itemize}
\item $S\subset \R^2$ is the {\em spatial reference object} of $\cal O$,
which is semi-algebraic\footnote{A semi-algebraic set in $\R^d$ is
a Boolean combination of sets of the form
$\{(x_1,x_2,\ldots,x_d)\mid p(x_1,x_2,\ldots,x_d)>0\}$, where $p$
is a polynomial with integer coefficients in the real variables
$x_1$, $x_2$, ..., $x_d$.} in $\R^2$;
\item $I\subset \R$ is the {\em time domain\/} of $\cal O$, which is a connected and bounded semi-algebraic set in $\R$ (i.e., a point or a bounded interval); and
\item $f:\R^2\times \R \rightarrow \R^2$ is the {\em transformation function} of $\cal
O$, which is semi-algebraic\footnote{A function $f$ is said to be
semi-algebraic if its graph is a semi-algebraic subset of
$(\R^2\times \R) \times\R^2$.} and continuous both in the time
coordinate and in the spatial coordinates.
\end{itemize}
The semantics of an atomic geometric object ${\cal O}=(S,I,f)$ is
the spatio-temporal object $st({\cal O})= \{(x,y;t)\in \R^2\times
\R\mid (\exists x')(\exists y')((x',y')\in S\lland t\in I\lland
(x,y)=f(x',y';t))\}.$ \qed
\end{definition}

We remark that this definition guarantees that there is a finite
representation of an atomic geometric object by means of the
polynomial inequalities that describe its reference object, its
time domain and the graph of its transformation function. This
means that this data model is within the {\em constraint model\/}
for databases (we refer to \cite{cdbook} for an overview of the
research results in this area).

\begin{definition}\rm
A {\em geometric object\/} is a finite set of atomic geometric
objects. The semantics of a geometric object $\{{\cal O}_1,\ldots,
{\cal O}_n\}$ is the union of the semantics of the atomic objects
that constitute it, i.e., the set $$\bigcup_{1\leq i\leq n}
st({\cal O}_i).$$
 \qed
\end{definition}

We agree that whenever we write ``the spatio-temporal object $\cal
O$'', where $\cal O$ is an (atomic) geometric object, we mean the
semantics of the (atomic) geometric object $\cal O$. Also, when
${\cal O}=(S,I,f)$ is an atomic object and $t\in I$, we will refer
to the set $\{(x,y)\mid (\exists x')(\exists y')((x',y')\in S\land
(x,y)=f(x',y',t) )\}$ as the {\em frame of  ${\cal O}$ at time
$t$\/} and we will denote it $f(S;t)$.

We define the {\em time domain\/} of a geometric object to be the
smallest time interval that contains all the time domains of the
composing atomic geometric objects. Recall that the smallest
interval containing a set of intervals is also known as the convex
closure of this set. We denote the convex closure of the sets
$I_1$, $I_2$, ..., $I_n$ by $\overline{\bigcup}_{i=1}^n I_i $.

Remark that a spatio-temporal object is empty (or non-existing)
outside the time domain of the geometric object that defines it.
Also, within its time domain a spatio-temporal object can be empty
(for instance, at any moment when no atomic geometric object
exists).

We conclude this section by remarking that the above introduced
notions of spatial and spatio-temporal object and of (atomic)
geometric object can be generalized to arbitrary dimension $d$ (by
simply substituting $d$ for $2$ in the above definitions). Since
all the results in this paper are formulated for dimension 2, we
have chosen not to use this generalization here.

\subsection{Practically relevant classes of geometric
objects}\label{classes}

Here, we define special classes of geometric objects that are
relevant to spatio-temporal database practice. These classes are
denoted by $$\langle{\cal S,F}\rangle$$ and they are determined by
the type $\cal S$ of spatial reference object and the type $\cal
F$ of transformation function. For clarity, a geometric object
belongs to a class if all of its atomic geometric objects belong
to that class.

The classes of geometric figures in the plane $\R^2$ that we will
consider are \begin{itemize}
\item \spoly, the class of arbitrary polygons,
\item \str, the class of arbitrary triangles, \item \strax, the class of triangles
with two sides parallel to the coordinate axes, and\item \srect\
the class of rectangles with all sides parallel to the coordinate
axes.
\end{itemize}

In this paper, we assume triangles, polygons and rectangles to be
filled objects. But since we allow two or more corner points of a
triangle or rectangle to coincide, the model can deal with
polylines and points too.  A line segment and a point are
considered triangles. Also line segments parallel to the axes and
points are considered rectangles. Finally, note that
$\ssrect\subset\sstrax\subset\sstr\subset\sspoly$.

The classes of transformation functions we will consider are
\begin{itemize} \item
\faff, the class of the affine transformations, \item \fsc, the
class of the  scalings,\item \ftrans, the class of the
translations, and
\item \fid, the class consisting of the identity mapping.\end{itemize}

It is clear that \fid, \ftrans\ and \fsc\ are subclasses of \faff.
 More technically, these classes are defined as follows.
The class \faff\ of affine transformations consists of the
mappings $\R^2\times \R \rightarrow \R^2$ of the form $$
(x,y;t)\mapsto \left(\!\begin{array}{@{}cc@{}}a(t) & b(t)\\ c(t) &
d(t)\end{array}\!\right) %\cdot
\left(\!\begin{array}{@{}c@{}}x
\\ y\end{array}\!\right) + \left(\!\begin{array}{@{}c@{}}e(t) \\f(t)
\end{array}\!\right),$$
where $a$, $b$, $c$, $d$, $e$, and $f$ are function from $\R$ to
$\R$ with $a(t)d(t)-c(t)b(t)\not = 0$ for all $t$ in the relevant
time domain.

The class \fsc\  of scalings consists of the affine
transformations for which the functions $b$ and $c$ are identical
to 0. The class \ftrans\ consists of the scalings for which the
functions  $a$ and $d$ are identical to 1.

For practical purposes we will only consider functions $a$, $b$,
$c$, $d$, $e$, and $f$ that are semi-algebraic and continuous as
required by the definition. These are \begin{itemize} \item the
rational functions (i.e., fractions of polynomial functions),
\item the polynomial functions and \item the linear polynomial
functions.\end{itemize}

The corresponding classes of transformations will be denoted using
superscripts ${\cal F}^{\rm Rat}$, ${\cal F}^{\rm Poly}$, and
${\cal F}^{\rm Lin}$. For example, \fscrat\ represents the class
of rational scalings. We assume that the time domain of an atomic
geometric object belongs to the domain of the transformation
function and that the denominator of a rational function in the
definition of a transformation is never zero in the closure of the
time domain (thus, the moving figure will remain within fixed
bounds during the time domain).

Note that the shape of a spatio-temporal object at a certain time
instant is not necessarily the same as the shape of the reference
object of the geometric object that gives rise to the
spatio-temporal object. For example, a rectangle is mapped to a
parallelogram under an affinity.

\subsection{Example}\label{vb1} Let ${\cal O}_A=(S_A,I_A,f_A)$ and ${\cal O}_B=(S_B,I_B,f_B)$ be two (atomic) geometric objects with spatial reference objects
$S_A$ and $S_B$ respectively the triangles with corner points
$(-1,0)$, $(1,0)$, $(0,1)$ and $(-1,0)$, $(1,0)$, $(0,-1)$, and
time domains $I_A=I_B=[0,2]$. In this time domain, $S_A$ remains
at its place (i.e., $f_A(x,y;t)=(x,y)$ for all $t$), while $S_B$
is translated with constant speed (equal to 1) in the direction of
the positive $y$-axis (i.e., $f_B(x,y;t)= (x,y+t)$. The functions
$f_A$ and $f_B$  belong to ${\cal F}^{\rm Lin}_{\rm Trans}$.

At $t = 0$ both objects intersect in a line segment. For $0<t<1$
they intersect in a hexagon, for $1\leq t <2$ in a quadrangle, and
finally for $t=2$ in a point.

\begin{figure}[h]
\centering
\input{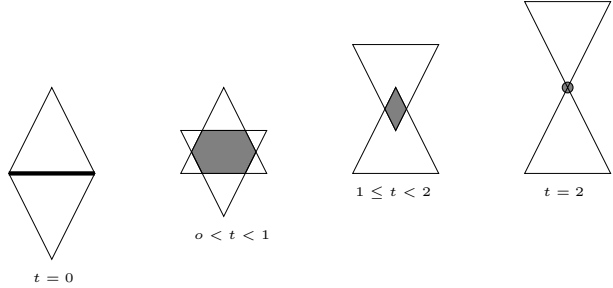} \caption{Two atomic geometric
objects. The time domain can be partitioned in four parts such
that the intersection of the two objects retains the same shape
during each element of the partition.}\label{fig1}
\end{figure}

\section{Closure properties under Boolean set operations}\label{results}

In this section, we work with the classes $\langle {\cal
S,F}\rangle$ introduced in the previous section, and we
investigate which of these classes $\langle {\cal S,F}\rangle$ are
closed under the Boolean set operations $\cup$ (union), $\cap$
(intersection) and $\setminus$ (set difference). We first define
what closure means.

\begin{definition}\label{closuredef} \rm
Let $\theta$ be one of the operations $\cup$, $\cap$ or
$\setminus$. We say that the class $\langle {\cal S,F}\rangle$ is
{\em (atomically) closed under $\theta$} if for any two (atomic)
geometric objects ${\cal O}_1$ and ${\cal O}_2$ in  $\langle {\cal
S,F}\rangle$ there exists a geometric object $\cal O$ in $\langle
{\cal S,F}\rangle$ such that $st({\cal O}) = st({\cal O}_1)
\mathrel{\theta} st({\cal O}_2)$. \qed
\end{definition}

We will refer to an object $\cal O$ that satisfies the condition
in the definition as an intersection, union or difference of
${\cal O}_1$ and ${\cal O}_2$ (they need not be unique).

For the union operation, the closure follows immediately from the
definition.

\begin{property}
For any class of objects $\cal S$ and any class of transformations
$\cal F$, $\langle {\cal S,F}\rangle$ is closed under $\cup$.\qed
\end{property}

For $\cap$ and $\setminus$ the situation is more complicated. The
next theorem is the main result that we want to prove in this
section. It summarizes the closure results for intersection and
set-difference.

\begin{theorem}\label{maintheorem}
For any class of objects $\cal S$ among \spoly, \str, \strax \/
and \srect \/ and any class of transformations $\cal F$ among
\faff, \fsc, \ftrans\/ and ${\cal F}_{\rm id}$, the closure with
respect to $\cap$ and $\setminus$ is summarized in the following
table.\\

\medskip

\noindent
 {\small
\begin{tabular}{|c|| c | c | c | c | c | c | c | c | c | c | }
\hline $\cap$, $\setminus$ & \faffrat & \faffpoly & \fafflin &
\fscrat & \fscpoly & \fsclin & \ftransrat & \ftranspoly &
\ftranslin & ${\cal F}_{\rm id}$\\ \hline\hline
 \spoly & $+$ & $-$ & $-$ & $-$ & $-$ & $-$ & $-$ & $-$ & $-$ & $+^\dagger$\\ \str &
$+$ & $-$ & $-$ & $-$ & $-$ & $-$ & $-$ & $-$ & $-$ &$+^\dagger$
\\
\strax  & $+$ & $-$ & $-$ &  $-$ &  $-$ & $-$ &  $-$ &  $-$ & $-$
&$-$\\ \srect & $+$ & $-$ & $-$ & $+$ & $+$ & $+$ & $-$ & $-$ &
$-$&$+$
\\
\hline
\end{tabular}
}
\\

\medskip\noindent
Closure is indicated by a $+$ sign, non-closure by a $-$ sign.\qed
\end{theorem}

The items marked with $\dagger$ are from~\cite{Wor94}.

The remainder of this section is devoted to proving this theorem.
We do this by first proving some lemmas in a first subsection that
reduce the number of cases that have to be looked at and by then
proving the remaining cases in a second subsection.

\subsection{Reduction properties}

The  properties in this section reduce the number of cases that
have to be investigated. First, we give a set-theoretic lemma that
will be used frequently.

\begin{lemma} \label{settheoretic}
Let $A_1,\ldots, A_n$ and  $B_1, \ldots,B_m$ be sets. Then
\begin{enumerate}
  \item[(a)] $(\bigcup_{i=1}^n A_i) \cap (\bigcup_{j=1}^m B_j) = \bigcup_{i=1}^n\bigcup_{j=1}^m
  (A_i\cap B_j)$,
  \item[(b)] $(\bigcup_{i=1}^n A_i) \setminus (\bigcup_{j=1}^m B_j) = \bigcup_{i=1}^n
  ((\cdots ((A_i\setminus B_1)\setminus B_2)\setminus \cdots  )\setminus B_m)$
\end{enumerate}

\end{lemma}

\smallskip
\par\noindent
{\bf Proof}. The first equality follows directly from
distributivity of intersection with respect to union.

The second equality can be proven by induction on $m$, using the
observation that $(\bigcup_{i=1}^n A_i) \setminus B =
(\bigcup_{i=1}^n A_i) \cap B^c= \bigcup_{i=1}^n (A_i \cap
B^c)=\bigcup_{i=1}^n (A_i \setminus B)$, where $B^c$ denotes the
complement of $B$ with respect to some universe. \qed

\medskip

The next property says that for $\cap$ and $\setminus$ closure and
closure on atomic objects coincide.

\begin{property}[Atomicity]\label{atomicity}
Let $\cal S$ be a class of objects and $\cal F$ a class of
transformations. Then
\begin{enumerate}
\item[(a)]
$\langle {\cal S,F}\rangle$ is closed under $\cap$ if and only if
it is atomically closed under $\cap$, and
\item[(b)] $\langle {\cal S,F}\rangle$ is closed under $\setminus$ if and only if it is atomically closed under
$\setminus$.
\end{enumerate}

\end{property}

\smallskip
\par\noindent
{\bf Proof}. Both for  (a) and  (b) the only-if direction is
obvious. So we concentrate on the if-direction.

For the if-direction of (a), assume that $\langle {\cal
S,F}\rangle$ is atomically closed under $\cap$ and let $\{{\cal
O}_{1,1}, {\cal O}_{1,2}, \ldots, \ab {\cal O}_{1,n}\}$ and
$\{{\cal O}_{2,1}, \ab {\cal O}_{2,2},\ab \ldots, {\cal
O}_{2,m}\}$ be two geometric objects from $\langle {\cal
S,F}\rangle$. By using Lemma~\ref{settheoretic} (a), we get
$$(\bigcup_{i=1}^n st({\cal O}_{1,i}) )\cap (\bigcup_{j=1}^m
st({\cal O}_{2,j}) )= \bigcup_{i=1}^n \bigcup_{j=1}^m (st({\cal
O}_{1,i}) \cap st({\cal O}_{2,j}) ).$$ Since $\cap$ is assumed to
be atomically closed, each $st({\cal O}_{1,i}) \cap st({\cal
O}_{2,j})$ can be written as a union $\bigcup_{k=1}^{l_{ij}}
st({\cal O}_{k,i,j})$, where each ${\cal O}_{k,i,j}$ is an atomic
geometric object. Therefore, the intersection of $\{{\cal
O}_{1,1}, {\cal O}_{1,2}, \ldots, {\cal O}_{1,n}\}$ and $\{{\cal
O}_{2,1}, {\cal O}_{2,2}, \ldots, {\cal O}_{2,m}\}$ can also be
written as $\bigcup_{i=1}^n \bigcup_{j=1}^m
\bigcup_{k=1}^{l_{i,j}} st({\cal O}_{k,i,j})$. This completes the
proof of the if-direction of (a).

For the if-direction of (b), assume that $\langle {\cal
S,F}\rangle$ is atomically closed under $\setminus$ and let
$\{{\cal O}_{1,1}, {\cal O}_{1,2}, \ab \ldots, \ab {\cal
O}_{1,n}\}$ and $\{{\cal O}_{2,1}, \ab {\cal O}_{2,2},\ab \ldots,
\ab {\cal O}_{2,m}\}$ be two geometric objects from $\langle {\cal
S,F}\rangle$. By using Lemma~\ref{settheoretic} (b), we get
$$(\bigcup_{i=1}^n st({\cal O}_{1,i}) )\setminus (\bigcup_{j=1}^m
st({\cal O}_{2,j}) )= \bigcup_{i=1}^n ((\cdots (( st({\cal
O}_{1,i})\setminus st({\cal O}_{2,1}))\setminus st({\cal
O}_{2,2}))\setminus \cdots ) \setminus st({\cal O}_{2,m})).$$

We prove, by induction on $m$, that $((\cdots (( st({\cal
O}_{1,i})\setminus st({\cal O}_{2,1}))\setminus st({\cal
O}_{2,2}))\setminus \cdots ) \setminus st({\cal O}_{2,m})) $ is of
the form $\bigcup_{k=1}^{l} st({\cal O}'_{k})$.  Since $\setminus$
is assumed to be atomically closed, $st({\cal O}_{1i}) \setminus
st({\cal O}_{21})$ can be written as a union
$\bigcup_{k=1}^{l_{1}} st({\cal O}'_{k})$, where each ${\cal
O}'_{k}$ is an atomic geometric object. This proves the case
$m=1$. Next, assume we have shown that $((\cdots (( st({\cal
O}_{1,i})\setminus st({\cal O}_{2,1}))\setminus st({\cal
O}_{2,2}))\setminus \cdots ) \setminus st({\cal O}_{2,m-1})) $ is
$\bigcup_{k=1}^{l} st({\cal O}'_{k})$ with all ${\cal O}'_k$
atomic geometric objects. Then $((\cdots (( st({\cal
O}_{1,i})\setminus st({\cal O}_{2,1}))\setminus st({\cal
O}_{2,2}))\setminus \cdots ) \setminus st({\cal O}_{2,m})) $ is
$(\bigcup_{k=1}^{l} st({\cal O}'_{k}))\setminus st({\cal
O}_{2,m})$, which is $\bigcup_{k=1}^{l} (st({\cal
O}'_{k})\setminus st({\cal O}_{2,m}))$, using
Lemma~\ref{settheoretic} (b). Again, since $\setminus$ is assumed
to be atomically closed, each of the sets $st({\cal
O}'_{k})\setminus st({\cal O}_{2,m})$ is of the form
$\bigcup_{r=1}^{l_k} st({\cal O}''_{r})$.
 Therefore, the set-difference of $\{{\cal O}_{1,1}, {\cal
 O}_{1,2},
\ldots, {\cal O}_{1,n}\}$ and $\{{\cal O}_{2,1}, {\cal O}_{2,2},
\ldots, {\cal O}_{2,m}\}$ is also the semantics of a geometric
object from $\langle {\cal S,F}\rangle$. This completes the proof.
\qed

\medskip

The following property states that intersection and set-difference
are equivalent with respect to closure.

\begin{property}[Equivalence of $\cap$ and $\setminus$]\label{equivalence} Let $\cal S$ be a
class of objects and $\cal F$ a class of transformations. Then
$\langle {\cal S,F}\rangle$ is closed under $\cap$ if and only if
it is closed under $\setminus$.

\end{property}

\smallskip
\par\noindent
{\bf Proof}. By Property~\ref{atomicity} it suffices to prove this
property for atomic geometric objects.

For the if-direction, assume that $\langle {\cal S,F}\rangle$ is
closed under $\setminus$ and let ${\cal O}_1$ and ${\cal O}_2$ be
two atomic geometric objects from $\langle {\cal S,F}\rangle$.
Since, $$st({\cal O}_1)\cap st({\cal O}_2)= (st({\cal O}_1)\cup
st({\cal O}_2))\setminus ((st({\cal O}_1)\setminus st({\cal
O}_2))\cup (st({\cal O}_2)\setminus st({\cal O}_1)))$$

and since $(st({\cal O}_1)\setminus st({\cal O}_2))$ and
$(st({\cal O}_2)\setminus st({\cal O}_1))$ are by assumption
$\bigcup_{i=1}^n st({\cal O}'_i)$ respectively $\bigcup_{j=1}^n
st({\cal O}''_j)$ with all ${\cal O}'_i$ and ${\cal O}''_j$ atomic
geometric objects. Therefore, $st({\cal O}_1)\cap st({\cal O}_2)$
equals $((\cdots (st({\cal O}_2)\setminus st({\cal
O}'_1))\setminus \cdots)\setminus st({\cal O}'_n ))\cup ((\cdots
(st({\cal O}_2)\setminus st({\cal O}''_1))\setminus
\cdots)\setminus st({\cal O}''_m ))$,  using
Lemma~\ref{settheoretic} (b). Using the argumentation from the
proof of the if-direction of (b) of Property~\ref{atomicity}, we
can show that this set is again a union of semantics of atomic
geometric objects from $\langle {\cal S,F}\rangle$.

For the only-if direction, assume that $\langle {\cal S,F}\rangle$
is closed under $\cap$ and let ${\cal O}_1=(S_1,I_1,f_1)$ and
${\cal O}_2=(S_2,I_2,f_2)$ be two atomic geometric objects from
$\langle {\cal S,F}\rangle$. We have to show that $st({\cal
O}_1)\setminus st({\cal O}_2)$ can be written as $\bigcup_{i=1}^n
st({\cal O}'_i)$, with ${\cal O}'_i$ atomic geometric objects. We
can restrict our attention to the set $st({\cal O}_1)\setminus
st({\cal O}_2)$ in the interval  $I_1\cap I_2$ rather than in the
complete interval $I_1\overline{\cup}I_2$ (since the
set-difference is empty in $I_2\setminus I_1$ and equal to ${\cal
O}_1$ in $I_1\setminus I_2$). Let $I$ denote the topological
closure of $I_1\cap I_2$. The set $S_B=\{(x,y)\in \R^2\mid
(\exists x')(\exists y')(\exists t)((x',y')\in S_1\lland t\in
I\lland f_1(x',y',t)=f_2(x,y;t))\}$ is compact (i.e.,
topologically closed and bounded) since it is the image of the
compact set $S_1\times I$ under the  continuous function
$f_2^{-1}\circ f_1$. Therefore, also $S=S_2\cup S_B$ is a compact
set in $\R^2$. Let $\alpha: (x,y)\mapsto (ax+ b_1, ay+b_2)$ be a
scaling followed by a translation that maps $S_2$ to a set that
strictly contains $S$ (this is possible since $S$ is bounded).
Remark that $\alpha$ maps any line to a parallel line. Let ${\cal
O}_3$ be the atomic geometric object $(\alpha(S_2), I, f_2)$. At
any moment $t$ in $I$, we thus have that $f_2(S_2;t)\subset
f_2(\alpha(S_2);t)$ (since affinities are monotone mappings) and
$f_1(S_1;t)\subset f_2(\alpha(S_2);t)$. Therefore, $st({\cal
O}_1)\setminus st({\cal O}_2)= st({\cal O}_1)\cap (st({\cal
O}_3)\setminus st({\cal O}_2))$.

\begin{figure}
\begin{center}
\centerline{\mbox{\psfig{file=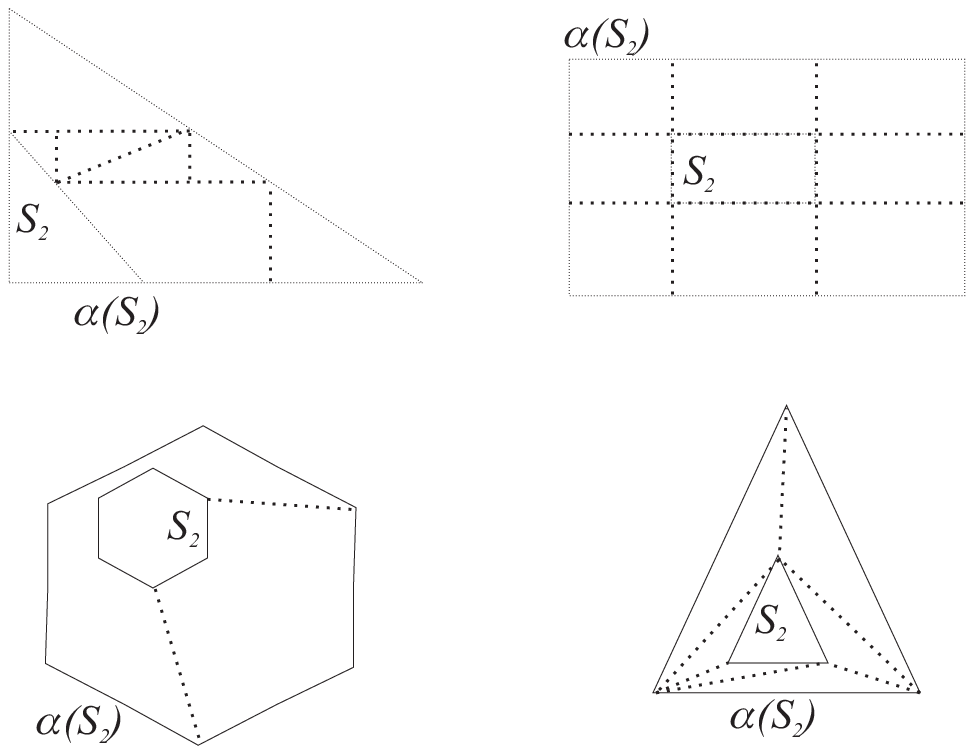}}} \caption[dummy]{Examples
of partitions of the set $\alpha(S_2)\setminus S_2$ for the
classes \spoly, \str, \strax, and \srect\ respectively.}
\label{polyfig}
\end{center}
\end{figure}

Now, $st({\cal O}_3)\setminus st({\cal O}_2)$ can always be
written as the semantics of a geometric object in $\langle {\cal
S,F}\rangle$ where $\cal S$ and $\cal F$ are any pairs allowed in
Theorem~\ref{maintheorem}. For each of the classes  \spoly, \str,
\strax, and \srect\ this is illustrated in Figure~\ref{polyfig}.
For each of these classes $\alpha(S_2)\setminus S_2$ can be
partitioned into a finite number of reference objects $T_1$, ...,
$T_n$ from these classes. So, define the atomic geometric objects
${\cal O}'_i=(T_i, I, f_2)$ ($1\leq i\leq n$). Then $st({\cal
O}_3)\setminus st({\cal O}_2)=\bigcup_{i=1}^n st({\cal O}'_i)$.
Therefore, $st({\cal O}_1)\cap (st({\cal O}_3)\setminus st({\cal
O}_2))= st({\cal O}_1)\cap \bigcup_{i=1}^n st({\cal O}'_i)=
\bigcup_{i=1}^n (st({\cal O}_1)\cap st({\cal O}'_i)) $. Since we
have assumed that $\langle {\cal S,F}\rangle$ is closed for
intersection,  the intersections $st({\cal O}_1)\cap st({\cal
O}'_i)$ can be written as $\bigcup_{k=1}^{l_i}st({\cal O}''_k)$
with ${\cal O}''_k$ atomic geometric objects from $\langle {\cal
S,F}\rangle$. Therefore also $st({\cal O}_1)\cap (st({\cal
O}_3)\setminus st({\cal O}_2))$ can be written as such a union.
This completes the proof.\qed

\medskip

A final reduction property says that the closure results for
polygons and triangles coincide. We can therefore concentrate on
triangles further on.

\begin{property}\label{polytotriangle}
Let $\cal F$ be a class of transformations, and let $\theta$ be
one of the operations $\cup$, $\cap$ or $\setminus$. Then $\langle
\sspoly , {\cal F}\rangle$ is closed under $\theta$ if and only if
$\langle \sstr , {\cal F}\rangle$ is closed under $\theta$.
\end{property}

\smallskip
\par\noindent
{\bf Proof}. This property follows from the fact that any atomic
geometric object ${\cal O}=(S,I,f)$ from  $\langle \sspoly , {\cal
F}\rangle$ corresponds to a geometric object from $\langle \sstr ,
{\cal F}\rangle$. Indeed, let $T_1,\ldots , T_n$ be an arbitrary
triangulation  of the polygon $S$. The geometric object $\{ {\cal
O}_1, \ldots , {\cal O}_n\}$ with ${\cal O}_i=(T_i,I,f)$ ($1\leq
i\leq n$) has the same semantics as ${\cal O}=(S,I,f)$.

So, if $\langle \sspoly , {\cal F}\rangle$ is closed under
$\theta$, then also any union, intersection or set-difference of
two elements of $\langle \sstr , {\cal F}\rangle$ is again a
geometric object of $\langle \sspoly , {\cal F}\rangle$ and
because of the above argument also of $\langle \sstr , {\cal
F}\rangle$.

On the other hand, suppose that $\langle \sstr , {\cal F}\rangle$
is closed under $\theta$. If ${\cal O}_1$ and ${\cal O}_2$ are
objects in $\langle \sspoly , {\cal F}\rangle$, then so are their
union, intersection or set-difference, since they are in  $\langle
\sstr , {\cal F}\rangle$, which is a subclass of $\langle \sspoly
, {\cal F}\rangle$. \qed

\subsection{Closure and non-closure proofs}

In this section, we complete the proof of
Theorem~\ref{maintheorem}, by means of a series of lemmas that
cover all the cases presented in the matrix of
Theorem~\ref{maintheorem}. Here, we take the reduction results of
the previous section into account. In particular, we only consider
intersections {\em or\/} set-differences of {\em atomic\/}
geometric objects, and we do not have to consider polygons any
more.

\subsubsection{Finite time partition}

Before giving these lemmas we introduce the technical notion of
{\em finite time partition}. This will be of use in many of the
proofs in this section. The finite time partition property tells
us  how and when the form (or appearance) of the intersection or
set-difference of two atomic geometric objects changes. We observe
that the intersection of two moving triangles can be empty, a
single point, a straight line segment, a triangle, a quadrangle, a
pentagon and a hexagon. The intersection of two moving rectangles
can be empty, a single point, a line segment or a rectangle. We
refer to all these different forms of the intersection or the
set-difference as their possible {\em shapes}. Also the difference
of two triangles or two rectangles can take a finite number of
different shapes. In the example in Figure~\ref{fig1}, the
intersection takes four different shapes, whereas the difference
takes five different shapes.

We define this notion now more technically. Let ${\cal
O}_1=(S_1,I_1,f_1)$ and ${\cal O}_2=(S_2,I_2,f_2)$ be two atomic
geometric objects with rational affine transformations with time
domains $I_1$ and $I_2$. In the following, we denote by $I_1
\mathbin{\overline \cup} I_2$ the convex closure of the set
$I_1\cup I_2$ in $\R$. Let $t$ be in $I_1 \mathbin{\overline \cup}
I_2$. Firstly, we call any line that intersects the border of
$f_i(S_i;t)$ in infinitely many points, a {\em carrier of the
frame $f_i(S_i;t)$} and denote it $car(f_i(S_i;t))$  ($i=1,2$).

\begin{definition}[Finite time partition]\label{def-finitetimepartition}
\rm We call a {\em finite time partition of ${\cal O}_1$ and
${\cal O}_2$} any partition of the interval $I_1
\mathbin{\overline \cup} I_2$ into a finite number of time
intervals $J_1,\ldots, J_m$ such that for any $t,t'\in J_i$ (and
all $1\leq i\leq m$), $car(f_1(S_1;t))\cup car(f_2(S_2;t))$ and
$car(f_1(S_1;t'))\cup car(f_2(S_2;t'))$ are topologically
equivalent sets\footnote{We call two subsets $A$ and $B$ of $\R^2$
{\em topologically equivalent\/} when there exists an
orientation-preserving homeomorphism $h$ of $\R^2$ such that
$h(A)=B$.} in $\R^2$. \qed
\end{definition}

\begin{property}\label{finitetimepartition}
Let ${\cal O}_1$ and ${\cal O}_2$ be two atomic geometric objects
with rational affine transformations with time domains $I_1$ and
$I_2$. There exists a finite time partition of ${\cal O}_1$ and
${\cal O}_2$.
\end{property}

\smallskip
\par\noindent
{\bf Proof.} Let ${\cal O}_1=(S_1,I_1,f_1)$ and ${\cal
O}_2=(S_2,I_2,f_2)$ be two atomic geometric objects satisfying the
conditions of the statement of this property. From the assumption
that the reference objects $S_1$ and $S_2$
 are semi-algebraic and the
transformation functions $f_1$ and $f_2$ are affine rational
functions, it follows that the sets $st({\cal O}_1)$ and $st({\cal
O}_2)$ are semi-algebraic subsets of $\R^{2}\times \R$ (for
details on this type of basic results on semi-algebraic sets, we
refer to Chapter 2 of~\cite{bcr}). Let $I$ be the set $I_1
\mathbin{\overline \cup} I_2$.

Also, the set $A=\bigcup_{t\in I_1 \mathbin{\overline \cup} I_2}
(car(f_1(S_1;t))\cup car(f_2(S_2;t)))$ is semi-algebraic, since it
can be defined in the first-order logic of the reals over the
semi-algebraic sets $st({\cal O}_1)$ and $st({\cal O}_2)$ (this
closure property of first-order logic over the reals can be found
in Chapter 2 of \cite{cdbook}). We can therefore consider the set
$A$ as a subset of $\R^{2}\times \R$ parameterized by the time
parameter $t$. It follows from {\em Semi-algebraic Triviality}
(Theorem 9.3.2 in~\cite{bcr} and also page 147
in~\cite{vandendries}) that the set $A$ induces a finite partition
on $I_1 \mathbin{\overline \cup}I_2$ such that in each partition
class $A$ remains topologically equivalent.\qed

\medskip

\subsubsection{Technical lemmas}

The following two lemmas are technical lemmas that say that
two/three points that move with their respective rational
affinities can be combined into one line/triangle that moves by a
single rational affinity. For the proofs we refer to the Appendix.

\begin{lemma}\label{polyaffineplus3}
Let ${\cal O}_i=(\{(x_i,y_i)\},I,g_i)$ ($i=1,2,3$) be three atomic
geometric objects with $g_i\in {\cal F}_{\rm Aff}^{\rm Rat}$. If
the three points $ (x_1,y_1)$, $(x_2,y_2)$ and $(x_3,y_3)$ form a
triangle $S$ (i.e., are not collinear) and if $g_1(x_1,y_1;t)$,
$g_2(x_2,y_2;t)$ and $g_3(x_3,y_3;t)$ form a triangle $S_t$ at any
moment $t\in I$ (i.e., are not collinear), then there exists an
atomic geometric ${\cal O}=(S,I,g)$ with $g\in {\cal F}_{\rm
Aff}^{\rm Rat}$ such that $g_i(x_i,y_i;t)=g(x_i,y_i;t)$ for all
$t\in I$ and $i=1,2,3$.\qed
\end{lemma}

\medskip

\begin{lemma}\label{polyaffineplus2}
Let ${\cal O}_i=(\{(x_i,y_i)\},I,g_i)$ ($i=1,2$) be two atomic
geometric objects with $g_i\in {\cal F}_{\rm Aff}^{\rm Rat}$. If
the two points $ (x_1,y_1)$ and $(x_2,y_2)$ form a line segment
$L$ (i.e., are not equal) and if $g_1(x_1,y_1;t)$ and
$g_2(x_2,y_2;t)$ form a line segment $L_t$ at any moment $t\in I$
(i.e., are not equal), then there exists an atomic geometric
${\cal O}=(L,I,g)$ with $g\in {\cal F}_{\rm Aff}^{\rm Rat}$ such
that $g_i(x_i,y_i;t)=g(x_i,y_i;t)$ for all $t\in I$ and $i=1,2$.
\qed
\end{lemma}

\medskip

 The next lemma shows that if two lines that move with a
rational affinity intersect, also the intersection point is moved
by a rational affinity. The proof of this lemma is in the
Appendix.

\begin{lemma}\label{intersection2lines}
Let ${\cal O}_i=(L_i,I,g_i)$ ($i=1,2$) be two atomic geometric
objects with $L_i$ line segments and $g_i\in {\cal F}_{\rm
Aff}^{\rm Rat}$. If the line segments $g_1(L_1;t)$ and
$g_2(L_2;t)$ intersect at any moment $t\in I$, then there exists
an atomic geometric ${\cal O}=(\{(x_0,y_0)\},I,g)$ with $g\in
{\cal F}_{\rm Aff}^{\rm Rat}$ that describes the intersection
point of $g_1(L_1;t)$ and $g_2(L_2;t)$ in $I$. \qed
\end{lemma}

\medskip

\subsubsection{Results for affinities}
 We can now start our series of closure and
non-closure lemmas and start with the affine transformations. For
the most general classes we have the following positive result.

\begin{lemma}\label{polyaffineplus}
The classes $\langle$\spoly,\faffrat$\rangle $ and $\langle$\str,
\faffrat$\rangle $ are closed under $\cap$ and $\setminus$.
\end{lemma}

\smallskip\par\noindent
{\bf Proof.} By Property~\ref{polytotriangle}, it suffices to show
this lemma for triangles. By Properties~\ref{atomicity} ({\em
atomicity\/}) and \ref{equivalence}, it suffices to show that the
intersection of two atomic geometric objects ${\cal
O}_1=(T_1,I_1,f_1)$ and ${\cal O}_2=(T_2,I_2,f_2)$ from
$\langle$\str, \faffrat$\rangle $ is represented by an object in
$\langle$\str, \faffrat$\rangle $.

According to Property~\ref{finitetimepartition} ({\em finite time
partition\/}), the intersection of the two moving triangles can
only take a finite number of different shapes, with each new shape
occurring in an element  of a finite partition of $I_1
\mathbin{\overline{\cup}} I_2$ into intervals $J_1,\ldots, J_m$
(in fact, we only have to consider $I_1\cap I_2$ here, since
outside this intersection the intersection of ${\cal O}_1$ and
${\cal O}_2$ is empty anyway). Let $J_l$ be an interval in this
partition. The intersection of ${\cal O}_1$ and ${\cal O}_2$ can
be a convex polygon (with at most six corner points), a line
segment or a single point in $J_l$.

First, suppose the intersection is a convex polygon. Let $t_0$ be
a point in $J_l$ (even if it is a degenerated interval, $J_l$
contains at least one point). We take the intersection of
$f_1(T_1;t_0)$ and $f_2(T_2;t_0)$ as reference object $P$. The set
$P\subset \R^2$ can be triangulated, for instance by connecting
its corner points to its point of gravity: this yields triangles
$T'_1,\ldots, T'_m$ (with $1\leq m\leq 6$). Each of the corner
points $(x_1,y_1)$, $(x_2,y_2)$, $(x_3,y_3)$ of a triangle $T'_j$
is moved in the time interval $J_l$ by a rational affinity (in
particular it is moved $f_1$ or $f_2$ applied to
 the inverse image of $f_1(\cdot;t_0)$, respectively
 $f_2(\cdot,t_0)$).
More specifically, a corner point of $T'_j$ is moved by $f_1$ if
it is originating from a corner point of ${\cal O}_1$;  a corner
point of $T'_j$ is moved by $f_2$ if it is originating from a
corner point of ${\cal O}_2$; Lemma~\ref{intersection2lines} shows
that there exists a rational affinity that moves a corner point of
$T'_j$ if it is an intersection point of side lines of ${\cal
O}_1$ and ${\cal O}_2$; a corner point of $T'_j$ can be taken to
be moved by $f_1$ if it is originating from the point of gravity
of $P$. Therefore, all corner points of $T'_j$ are moved by a
rational affinity. Lemma~\ref{polyaffineplus3} guarantees the
existence of a rational affinity $f_j$ that moves $T'_j$. The
intersection of ${\cal O}_1$ and ${\cal O}_2$ in $J_l$ is
therefore described by the atomic geometric objects
$(T'_j,J_l,f_j)$ ($1\leq j\leq m\leq 6$).

Second, we investigate the situation if the intersection of ${\cal
O}_1$ and ${\cal O}_2$ is a line segment. The end points of the
intersection originate from ${\cal O}_1$ or  ${\cal O}_2$ or can
be the result of intersecting side lines of ${\cal O}_1$ and
${\cal O}_2$. In both cases, (from Lemma~\ref{intersection2lines}
for an intersection point) it is clear that the two end points are
moved by  a rational affine transformation.
Lemma~\ref{polyaffineplus2} then shows that there exists a single
rational affine transformation $f$ to move the intersection. This
intersection can therefore be described by an atomic geometric
object $(L,J_l,f)$, where $L$ is some line segment.

Third, we look at the case where the intersection is a single
point. This point can originate from ${\cal O}_1$ or ${\cal O}_2$
or can be the result of intersecting side lines of ${\cal O}_1$
and ${\cal O}_2$. In both cases, (from
Lemma~\ref{intersection2lines} for an
 intersection point), it is
clear that in this case the intersection's movement is a rational
affine transformation.
 \qed

\medskip\par\noindent In general, if the affine transformations of
${\cal O}_1$ and ${\cal O}_2$ are given by polynomial or linear
functions, the corner points $(x_1,y_1)$, $(x_2,y_2)$ and
$(x_3,y_3)$ of triangles in the intersection (or difference) are
in general rational in these functions. The computations in the
proof of the Lemmas~\ref{polyaffineplus3}, \ref{polyaffineplus2}
and \ref{intersection2lines} suggest that this leads to
non-closure.

\begin{lemma}\label{polyaffinemin}
The classes $\langle$\spoly,\faffpoly$\rangle $,
$\langle$\spoly,\fafflin$\rangle $,
$\langle$\str,\faffpoly$\rangle$ and $\langle$\str,
\fafflin$\rangle $ are not closed under $\cap$ and $\setminus$.
\end{lemma}

\smallskip\par\noindent
{\bf Proof.} It suffices to prove the lemma for triangles. We give
a counterexample for intersection that serves for both classes
$\langle$\str, \fafflin$\rangle $ and $\langle$\str,
\faffpoly$\rangle $. Consider two atomic geometric objects ${\cal
O}_1$ and ${\cal O}_2$ with reference objects triangles with
corner points $(1,1)$, $(3,1)$, $(2,3)$ and $(2,2)$, $(4,2)$,
$(3,4)$, respectively. The affine transformations of these
triangles are given by the matrices

\begin{center}
\begin{tabular}{ccc}$\left(
\begin{array}{@{}c c@{}} t & 2t\\ 3t &
t\end{array}\right)$ &{\rm and\ } &$\left( \begin{array}{@{}c
c@{}}t & 2t+1\\ t & 3t+1\end{array}\right),$
\end{tabular}\end{center}

\noindent respectively. Assume these objects are moved in some
interval of the strictly positive $t$-axis (for example
$I=[1,2]$), the intersection of the two objects is a triangle with
corner points $(6t+2,8t+2)$,
$(\frac{1}{2}t\frac{(181t+70)}{(13t+4)},\frac{1}{2}t\frac{(243t+70)}{(13t+4)})$
and $(\frac{29}{4}t,\frac{37}{4}t)$.

Assume that  this triangle could be represented as a geometric
object $\{{\cal O}_1,\ldots, {\cal O}_m\}$ from $\langle$\str,
\faffpoly$\rangle $. Then, there exists some subinterval $J$ of
$I$ during which the corner point
$(\frac{1}{2}t\frac{(181t+70)}{(13t+4)},\frac{1}{2}t\frac{(243t+70)}{(13t+4)})$
is the image of a corner point $(x_0,y_0)$ of a reference triangle
that is transformed by a polynomial (or linear) affinity. We
therefore have that, for instance the $x$-coordinate
$\frac{1}{2}t\frac{(181t+70)}{(13t+4)}$ of the above point is of
the form $a(t)x_0+b(t)y_0+e(t)$  for $t\in J$ with $a(t)$, $b(t)$
and $e(t)$  polynomials (or linear polynomials) in $t$. Therefore,
$181t^2+70t-2(a(t)x_0+b(t)y_0+e(t))(13t+4)=0$ for all $t\in J$.
Since the number of zero's of this polynomial exceeds its degree,
it is identical to zero. Therefore, $a(t)x_0+b(t)y_0+e(t)$ is of
the form $\alpha t+\beta$. This leads to the conditions $\beta
=0$, $181=26\alpha$ and $70=8\alpha$. There is no solution and we
have a contradiction. \qed

\medskip

\begin{lemma}\label{rectaffineplus}
The classes  $\langle$\srect,\faffrat$\rangle $ and
$\langle$\strax,\faffrat$\rangle $ are closed under $\cap$ and
$\setminus$.
\end{lemma}

\smallskip

\smallskip\par\noindent
{\bf Proof.}  Let us first consider the class
$\langle$\srect,\faffrat$\rangle $. Because of
Lemmas~\ref{atomicity}, and~\ref{equivalence}, it suffices to
consider the intersection of two atomic geometric objects ${\cal
O}_1=(R_1,I_1,f_1)$ and ${\cal O}_2=(R_2,I_2,f_2)$. The image of a
rectangle under an affinity is a parallelogram. The shape of the
intersection of $f_1(R_1;t)$ and $f_2(R_2;t)$ for some $t$ in
$I_1\cap I_2$ can therefore be a convex polygon with at most eight
corner points, a line segment or a point.

In any of these cases, we can copy the argumentation used in the
proof of Lemma~\ref{polyaffineplus}. In case the intersection is a
line segment or a point, this settles the case. In the case where
it is a convex polygon, we can reuse the triangulation technique
presented in the proof of Lemma~\ref{polyaffineplus}, now noting
that it can consist of at most eight triangles instead of six. So,
we get that the intersection of ${\cal O}_1$ and ${\cal O}_2$ can
be described by the atomic geometric objects $(T'_j,J_l,f_j)$
($1\leq j\leq m\leq 8$), where the $T'_j$ are triangles and the
$f_j$ are rational affinities.

For the purpose of this lemma, we need to describe the
intersection of ${\cal O}_1$ and ${\cal O}_2$ by means of moving
rectangles, however. This can be achieved by replacing each of the
triangles $T'_j$ by three rectangles $R_{1j}$, $R_{2j}$ and
$R_{3j}$. Let the corner points of $T'_j$ be $ (x_1,y_1)$,
$(x_2,y_2)$ and $(x_3,y_3)$. The rectangle $R_{ij}$ are chosen
such that a constant affinity $f_{ij}$ maps $R_{ij}$ to the
parallelogram with corner points  $ (x_i,y_i)$,
$\frac{1}{2}(x_1+x_2,y_1+y_2)$, $\frac{1}{2}(x_1+x_3,y_1+y_3)$ and
$\frac{1}{2}(x_3+x_2,y_3+y_2)$ ($i=1,2,3$). So, $T'_j$ is the
union of the three parallelograms: $T'_j=f_{1j}(R_{1j})\cup
f_{2j}(R_{2j})\cup f_{3j}(R_{3j})$.

So, if we replace $(T'_j,J_i,f_j)$ by $(R_{ij},J_l, f_j \circ
f_{ij})$ we get a description of the intersection of ${\cal O}_1$
and ${\cal O}_2$ during $J_l$ in terms of atomic geometric objects
from $\langle$\srect,\faffrat$\rangle $.

The closure result for  $\langle$\strax,\faffrat$\rangle $
 can be obtained by further dividing the rectangles $R_{i,j}$ along a diagonal
 into two triangles from \strax.\qed

\medskip

The following lemma concludes the results for affinities.

\smallskip

\begin{lemma}\label{rectaffinemin}
The classes $\langle$\srect,\faffl$\rangle $ and
$\langle$\strax,\faffl$\rangle $ are not closed under $\cap$ and
$\setminus$ for ${\rm L}\in \{{\rm Lin}, {\rm Poly}\}$.
\end{lemma}

\smallskip

\par\noindent {\bf Proof.}
First, let us look at $\langle$\srect,\faffl$\rangle $.  We give a
counterexample for intersection that serves for both classes
$\langle$\srect, \fafflin$\rangle $ and $\langle$\srect,
\faffpoly$\rangle $. We modify the counterexample from the proof
of Lemma~\ref{polyaffinemin}. Consider two atomic geometric
objects ${\cal O}_1$ and ${\cal O}_2$ with reference objects
rectangles with corner points $(1,1)$, $(3,1)$, $(1,3)$, $(3,3)$
and $(2,2)$, $(4,2)$, $(2,4)$, $(4,4)$, respectively. The affine
transformations of the rectangles are given by the matrices

\begin{center}
\begin{tabular}{ccc}$\left(
\begin{array}{@{}c c@{}} t & 2t\\ 3t &
t\end{array}\right)$ &{\rm and\ } &$\left( \begin{array}{@{}c
c@{}}t & 2t+1\\ t & 3t+1\end{array}\right),$
\end{tabular}\end{center}

\noindent respectively.

In some interval of the strictly positive $t$-axis, the
intersection of the two objects is a triangle with corner points
$(6t+2,8t+2)$,
$(t\frac{(28t+15)}{(3t+2)},3t\frac{(13t+2)}{(3t+2)})$ and
$(\frac{21}{2}t,\frac{17}{2}t)$.

The same type of argumentation as in the proof of
Lemma~\ref{polyaffinemin}, can be  used to show that at least a
rational affinity is needed to describe the intersection.
Therefore, both $\langle$\srect,\fafflin$\rangle$ and
$\langle$\srect,\faffpoly$\rangle$ are not closed for intersection
and set-difference.

Secondly, for $\langle$\strax,\faffl$\rangle $, we can reuse the
above counterexample leaving out the corner points $(1,1)$ and
$(4,4)$ respectively. The intersection remains the same and the
argumentation can be repeated. \qed

\medskip

The proof of Lemma~\ref{polyaffineplus} is based on the property
that affinities do not preserve parallelism to the axes. We will
see later that for scalings, which do preserve parallelism to the
axes, the class of the objects of \strax\ is not closed.

\smallskip

\subsubsection{Results for scalings}

We divide the results for scalings into one positive and two
negative results.

\begin{lemma}\label{rectscalings}
$\langle$\srect, \fscl $\rangle $ is closed under $\cap$ and
$\setminus$ for ${\rm L}\in \{{\rm Lin}, {\rm Poly}, {\rm Rat}\}$.
\end{lemma}

\par\noindent{\bf Proof.}
Because of Lemmas~\ref{atomicity}, and~\ref{equivalence}, it
suffices to consider the intersection of two atomic geometric
objects ${\cal O}_1=(R_1,I_1,g_1)$ and ${\cal O}_2=(R_2,I_2,g_2)$.

According to Property~\ref{finitetimepartition}, the intersection
of the two rectangles takes different shapes in elements of a
finite partition of $I_1\cap I_2$ (we only consider this
intersection, since elsewhere in  $I_1\mathbin{\overline{\cup}}
I_2$ the intersection of ${\cal O}_1$ and ${\cal O}_2$ is empty in
any case). Let $J$ be an interval in this partition. First, we
remark that scalings map lines that are parallel to the $x$-axis
or to the $y$-axis to a parallel line. Therefore, at any moment
$t$ in $J$ both the frame of ${\cal O}_1$ and the frame of ${\cal
O}_2$ are rectangles with sides parallel to the coordinate axis.

Let us assume that the intersection of ${\cal O}_1$ and  ${\cal
O}_2$ is a rectangle in $J$.

We remark that this intersection rectangle is uniquely determined
by the coordinates of its upper-left corner point $(x_{ul}(t),
y_{ul}(t))$ and the coordinates of the lower-right corner point
$(x_{lr}(t), y_{lr}(t))$. Let assume the upper-left corner point
of the intersection comes from ${\cal O}_1$ and the lower-right
from ${\cal O}_2$ (possibly we have to work with the upper-right
and lower-left corners, but this is equivalent). Let the scaling
of ${\cal O}_1$ be determined by $a_1(t)$, $b_1(t)$, $e_1(t)$,
$f_1(t)$ and the one of ${\cal O}_2$ by $a_2(t)$, $b_2(t)$,
$e_2(t)$, $f_2(t)$ (following the matrix notation of
section~\ref{classes}).

The intersection is an atomic geometric object ${\cal O}=(R, J,
f)$ composed as follows. The reference rectangle $R$ has as
upper-left corner point $(x_{ul}, y_{ul})$ the upper-left corner
point of the reference object $R_1$ of ${\cal O}_1$ and as
lower-right corner point $(x_{lr}, y_{lr})$ the lower-right corner
point of the reference object $R_2$ of ${\cal O}_2$ (if $(x_{lr},
y_{lr})$ and $(x_{lr}, y_{lr})$ have an $x$- or $y$-coordinate in
common, we work with $(x_{lr}+1, y_{lr}+1)$ instead of $(x_{lr},
y_{lr})$ and replace $e_2(t)$ with $e_2(t)-a_2(t)$ and $f_2(t)$
with $f_2(t)-b_2(t)$ in the description of $g_2$). The
transformation function $g$ of ${\cal O}$ is determined
  by
\\
\begin{center}
\begin{tabular}{r c l}
$a(t)$ & = &$\frac{(a_1(t)x_{ul}-a_2(t)x_{lr}
+e_1(t)-e_2(t)}{x_{ul}-x_{lr}},$\\

$b(t)$ & = &$\frac{(b_1(t)y_{ul}-b_2(t)y_{lr}
+f_1(t)-f_2(t)}{y_{ul}-y_{lr}},$\\

$e(t)$ & = &$\frac{((a_2(t)-a_1(t))x_{ul}x_{lr}-e_1(t)x_{lr}
+e_2(t)x_{ul}}{x_{ul}-x_{lr}},$\\

$f(t)$ & = &$\frac{((b_2(t)-b_1(t))y_{ul}y_{lr}-f_1(t)y_{lr}
+f_2(t)y_{ul}}{y_{ul}-y_{lr}}.$\\
\end{tabular}
\end{center}

 These formulas show that
if the transformations of ${\cal O}_1$ and ${\cal O}_2$ are
rational, polynomial, respectively linear, then also $a(t)$,
$b(t)$, $e(t)$, $f(t)$ are rational, polynomial, respectively
linear.

The cases where  the intersection of ${\cal O}_1$ and  ${\cal
O}_2$ is a line segment or point in $J$ are analogous to but
simpler than the previous case. \qed

\begin{figure}[h]
\centering
\input{fig4.pstex_t}
\caption{Counterexamples for intersection (A) and difference (B)
for the classes $\langle$\strax$,$\fsclin$\rangle$} \label{fig2}
\end{figure}

\begin{lemma}\label{rechthoekigescalings}
The classes $\langle $\strax$, $\fscl$\rangle$ and  $\langle
$\strax$, \{{\rm id}\}\rangle$ are not closed under $\cap$ and
$\setminus$ for ${\rm L}\in \{{\rm Lin}, {\rm Poly}, {\rm Rat}\}$.
\end{lemma}

\smallskip
\par\noindent
{\bf Proof}. Consider the triangle with corner points $(0,0)$,
$(1,0)$ and $(0,1)$ and the triangle with corner points
$(\frac{1}{3},1)$, $(\frac{2}{3},1)$ and $(\frac{2}{3},0)$, both
transformed by the identity transformation. Their intersection
(for an illustration see (A) of Figure~\ref{fig2}) cannot be
described as a finite union of elements of $\langle $\strax$,
$\fscl$\rangle$ since scalings map lines that are parallel to a
coordinate axis to a parallel line. (Remember, for affinities,
this class was closed, partly because affinities do not
necessarily preserve parallelism with the coordinate axis.) \qed

\medskip
The following lemma could be left out since it is implied by
Lemma~\ref{driehoekscalings}. We give it since its proof is
conceptually easier, however.

\begin{lemma}\label{driehoekscpollin}
The classes $\langle $\str$, $\fscl$\rangle$ and
$\langle$\spoly,\fscl$\rangle$ are not closed under $\cap$ and
$\setminus$ for ${\rm L}\in \{{\rm Lin}, {\rm Poly}\}$.
\end{lemma}

\smallskip

\par\noindent
{\bf Proof}. Because of Properties~\ref{atomicity}
and~\ref{polytotriangle} it suffices to prove this for atomic
geometric objects that have a triangle as a reference object.
Consider the triangle with corner points $(0,0)$, $(0,1)$ and
$(1,0)$, and the triangle with corner points $(0,0)$, $(1,1)$ and
$(1,0)$. Their respective transformation functions are the
scalings

\begin{center}
\begin{tabular}{lll}$\left(
\begin{array}{@{}c c@{}} 1 & 0\\ 0 &
t+1\end{array}\right)$ & and &$\left( \begin{array}{@{}c c@{}}1 &
0\\ 0 & 2t+1\end{array}\right).$\end{tabular}
\end{center}
 We consider for both
objects the time interval $[0,5]$. At any moment during this
interval the intersection is given by the triangle with corner
points $(0,0)$, $(\frac{t+1}{3t+2}, \frac{(t+1)(2t+1)}{3t+2})$ and
$(1,0)$. Assume that this intersection is described by a geometric
object $\{ {\cal O}_1,\ab \ldots, \ab {\cal O}_m\}$ from $\langle
$ \str, \fscl$\rangle$. At least one of the atomic objects
describes a moving triangle  that contains  $(\frac{t+1}{3t+2},
\frac{(t+1)(2t+1)}{3t+2})$ as a corner point during some
subinterval of $[0,5]$.  The $x$-coordinate $\frac{t+1}{3t+2}$ is
therefore of the form $a(t)x_0+e(t)$ with $x_0$ the $x$-coordinate
of some corner point of a reference object, and $a(t)$ and $e(t)$
functions appearing in its transformation matrix. Therefore,
$a(t)x_0+e(t)$ has degree $0$, i.e., it is a number, say $\alpha$.
But then $\alpha(3t+2)$ and $t+1$ should be identical polynomials,
leading to the equations $3\alpha =1$ and $2\alpha=1$ that clearly
do not have a solution. It can therefore not be a linear or
polynomial transformation. \qed

\medskip

The next lemma completes the proofs for scalings.

\begin{lemma}\label{driehoekscalings}
Neither $\langle $\str$, $\fscrat$\rangle$ nor
$\langle$\spoly,\fscrat$\rangle$ is closed under $\cap$ and
$\setminus$.
\end{lemma}

\smallskip

\par\noindent
{\bf Proof}. Because of Properties~\ref{atomicity}
and~\ref{polytotriangle} it suffices to prove this lemma for
atomic geometric objects that have a triangle as a reference
object. We give an example of two atomic geometric objects ${\cal
O}_1$ and ${\cal O}_2$ that have an intersection that cannot be
described in $\langle $\str$, $\fscrat$\rangle$.

 Let the reference triangle of the
atomic geometric object ${\cal O}_1$ have corner points $(0,0)$,
$(1,1)$ and $(\frac{1}{3},\frac{1}{2})$ and let the transformation
of this object be the scaling that maps $(x,y)$ to $$\left(
\begin{array}{@{}c c@{}}-\frac{3(t+1)}{t+3} & 0\\ 0 & -(t+1)
\end{array} \right)\left(
\begin{array}{@{}c@{}}x
\\ y \end{array} \right) +  \left( \begin{array}{@{}c@{} }\frac{3(t+1)}{t+3}
\\ t+1 \end{array} \right).$$
Let the reference triangle of the atomic geometric object ${\cal
O}_2$ have corner points $(0,0)$, $(1,1)$ and $(0,1)$ and let the
scaling of this object be the time-independent mapping that maps
$(x,y)$ to $$\left(
\begin{array}{@{}c c@{}}4 & 0\\ 0 & 4
\end{array} \right)\left(
\begin{array}{@{}c@{}}x
\\ y \end{array} \right) +  \left( \begin{array}{@{}c@{} }-2
\\ -2 \end{array} \right).$$
We consider both objects in the time interval $(0,\frac{1}{2})$.
At any moment during this interval the intersection is given by
the triangle with corner points $(0,0)$, $(\frac{3(t+1)}{t+3},
t+1)$ and $(1,1)$. We remark that the point $(\frac{3(t+1)}{t+3},
t+1)$ is situated above the diagonal $y=x$ and that in the limit
towards 0, this point converges to $(1,1)$. In other words, the
intersection is always a triangle during the time interval
$(0,\frac{1}{2})$, but it converges to a line segment for $t$
going to 0. It is easily verified that this intersection cannot be
described as the image of a single triangle under a scaling from
\fscrat.

More generally, assume that this intersection is described by a
geometric object $\{ {\cal O}_1,\ab \ldots, \ab {\cal O}_m\}$ from
$\langle $ \str, \fscrat$\rangle$. At least one of the atomic
objects describes a moving triangle  that covers a line segment
connecting $(0,0)$ and $(f(t),f(t))$ of the line connecting
$(0,0)$ and $(1,1)$ during a time interval $(0,\varepsilon]$ with
$\varepsilon>0$ (without loss of generality this interval can be
assumed to be closed on the right side). Let the third cornerpoint
$(g(t),h(t))$ be situated in the interior of the intersection
triangle with cornerpoints $(0,0)$, $(\frac{3(t+1)}{t+3}, t+1)$
and $(1,1)$. Let the scaling of this object be the one that maps
$(x,y)$ to $$\left(
\begin{array}{@{}c c@{}}a(t) & 0\\ 0 & b(t)
\end{array} \right)\left(
\begin{array}{@{}c@{}}x
\\ y \end{array} \right) +  \left( \begin{array}{@{}c@{}} c(t)
\\ d(t) \end{array} \right),$$
where $(a(t)$, $b(t)$, $c(t)$ and $d(t)$ are rational functions of
$t$. Without loss of generality the reference triangle of this
atomic object can be assumed to have cornerpoints $(0,0)$,
$(1,1)$, and $(a,b)$, where the first is mapped to $(0,0)$, the
second to $(f(t),f(t))$ and the third to $(g(t),h(t))$. Since we
assume this reference object to be a triangle, we have $a\not=b$.
It then follows that $a(t)$ and $b(t)$ must be equal to $f(t)$ and
that $c(t)$ and $d(t)$ must be constant 0. Therefore, this scaling
maps the third cornerpoint $(a,b)$ to $(g(t),h(t))=(a f(t),b
f(t))$. Both $a$ and $b$ are therefore strictly positive. Since
the point $(a f(t),b f(t))$ is situated at the same side of the
diagonal $y=x$ as the point $(0,1)$, we get the condition $b
f(t)-a f(t)> 0$, or $b> a$. On the other hand, this point is
situated on the same side as $(0,1)$ of the line connecting
$(0,0)$ and $(\frac{3(t+1)}{t+3}, t+1)$. Therefore, we get $$
(t+1)a f(t)-\frac{3(t+1)}{t+3}b f(t)>0.$$ From this
$a-\frac{3}{t+3}b>0$ follows, or $\frac{3b}{a}<t+3$. Since
$t\mapsto t+3$ is strictly increasing in $(0,\varepsilon]$ and has
infimum 3 over this interval, we get $\frac{3b}{a}\leq 3$, or
$b\leq a$. This contradicts $b>a$, that we obtained before. This
concludes the proof. \qed

\smallskip

\subsection{Results for translations}

We give a general negative result for translations.

\begin{lemma}%\label{translatielemma}
For each of the classes $\cal S$ considered in the previous
section,  the class $\langle {\cal S},$ \ftransl$\rangle$ is not
closed under $\cap$ and $\setminus$, for ${\rm L}\in \{{\rm Lin},
{\rm Poly}, {\rm Rat}\}$.
\end{lemma}

\smallskip

\par\noindent
{\bf Proof.} First, we remark that translations preserve the shape
and area of objects and the length of lines.

Consider now two reference objects, located in the plane $t=0$,
from each of the relevant classes that have the interval $[0,1]$
on the $x$-axis as one of their sides. Let one reference object be
located above the $x$-axis and the second be located below the
$x$-axis. Let the first object undergo the translation
$(x,y)\mapsto (x-t,y)$ in the direction of the negative $x$-axis
and let the second object undergo the translation $(x,y)\mapsto
(x+t,y)$ in the opposite direction, both in the time interval
$[0,t_0]$, for some $t_0>0$.

Then it is clear that the intersection of these objects is a
shrinking line segment during the time interval $[0,t_0]$. So, in
any of the cases, the intersection cannot be described as a finite
union of translated objects. \qed

\subsection{Results for the identity}

For completeness, we also give the results for the identity
mapping.

\begin{lemma}%\label{translatielemma}
The classes $\langle$\spoly,\fid$\rangle$,
$\langle$\str,\fid$\rangle$ and $\langle$\srect,\fid$\rangle$ are
closed under $\cap$ and $\setminus$. The class
$\langle$\strax,\fid$\rangle$ is not closed under $\cap$ and
$\setminus$.
\end{lemma}

\smallskip

\par\noindent
{\bf Proof.} For the positive closure results, it suffices to
remark the following. The intersection of two polygons is again a
polygon (if line  segments and points are considered to be in this
class). The intersection of two triangles is a convex polygon with
at most six corner points that can be triangulated, i.e., written
as a disjoint union of triangles. The intersection of two
rectangles is a rectangle, a line segment parallel to a coordinate
axis, or a point.

For the negative result, we remark that the intersection of two
reference objects from \strax\ cannot necessarily be written as a
finite union of such objects. Figure~\ref{fig2} contains  an
example. \qed

\medskip Now we have proven all the closure and non-closure results listed in the
table of Theorem~\ref{maintheorem}.

\section{The extended data model}

It is clear that the model for representing spatio-temporal data,
that we have presented in Section~\ref{defs}, gives mostly
negative closure results (see Theorem~\ref{maintheorem}) for the
classes of objects we considered important for spatio-temporal
practice. The only classes that seem to be useful for further
investigation are $\langle{\cal S}$, \faffrat$\rangle$, for any of
the considered classes $\cal S$ of reference objects.

In this section, we will enrich the data model and get better
closure results. We will also study normal forms for objects in
this enriched model.

In Section~\ref{defs}, we defined a geometric object as a finite
union of atomic objects. We could now try to modify this
definition by allowing other operations than union in the
construction of geometric objects from atomic geometric objects.
The exhaustive list of alternative definitions that could be
considered are: a geometric object is obtained from atomic
geometric objects by means of
\begin{itemize}
\item[(a)] union (see Section~\ref{results});
\item[(b)] intersection;
\item[(c)] set-difference;
\item[(d)] intersection and set-difference; and finally
\item[(e)] union, intersection and set-difference.
\end{itemize}

In this paper, we will not investigate alternatives (b), (c) and
(d). These alternatives may be interesting from a mathematical
point of view, but in any practical application it is natural to
allow union in the construction of spatio-temporal objects. In
fact it is easy to see that for instance alternative (b) gives
even worse closure results. Hereto, we first make two basic
observations. Firstly, it is clear that the intersection of convex
objects always results in a convex object, and that the affine
transformation of a convex object remains a convex object.
Secondly, the intersection of connected convex objects is again
connected. It should be clear therefore that when the reference
objects are triangles or rectangles, then whenever a union has two
connected components, it cannot be written as an intersection of
atomic geometric objects.

For alternative (c), we remark that in contrast to the
intersection, the difference of two convex objects can result in a
non-convex object, or in a set of disjoint objects. So, it is
possible to describe a wide class of objects as the difference of
some atomic objects. But, this approach has two major drawbacks:

\begin{enumerate}
\item If we want to describe a certain object ${\cal O}$ as the difference of some other objects
${\cal O}_1 \ldots {\cal O}_k$, we have to artificially introduce
those objects ${\cal O}_1, \ldots, {\cal O}_k$ into the database.
There is no way of controlling the number of objects that have to
be introduced, as this depends on the exact shape of the object
${\cal O}$.
\item The difference operator is not associative, so in the worst case the depth of the tree
describing the relation between the objects equals the number of
objects. For practical applicability of our model, we should have
a tree with limited depth. (One way of achieving this is to define
a normal form, see further).
\end{enumerate}

Only alternative (e) will be further investigated here.

\subsection{The extended data model}

First, we define the extended model. Atomic geometric objects are
defined as in Section~\ref{defs}.

\begin{definition}[Extended data model]\label{extendeddatamodel}
\rm
An {\em extended geometric object\/} is a binary tree, where each
non-leaf node has two children, where each of the nodes is labeled
with $\cup$, $\cap$ or $\setminus$ and where each leaf is labeled
with an atomic geometric object.

The semantics of a geometric object is defined (recursively
starting from the root of the tree) as the semantics of its root.
If a node $n$ of the binary three has a left child $lc$ and a
right child $rc$, and if the root is labeled $\theta$ (with
$\theta \in \{\cup,\cap, \setminus\}$), the semantics $sem(n)$ of
node $n$ is by definition $sem(lc)\mathrel{\theta}sem(rc)$. The
semantics of a leaf labeled with the atomic geometric object
${\cal O}$ is $st({\cal O})$.\qed
\end{definition}

We define the {\em time domain\/} of an extended geometric object
to be the convex closure of the union of the time domains of all
the composing atomic geometric objects.

By slight abuse of notation, we will write down binary trees as in
Definition~\ref{extendeddatamodel} in the usual set-theoretic
notation. The expression ${\cal O}_1\cup ({\cal O}_2\cap ({\cal
O}_3\setminus {\cal O}_1) )$ is an example.

The following property is trivial and says that this model is
closed for all Boolean set operations.
\begin{property} For all the classes $\langle {\cal S},{\cal F}\rangle$ considered in Section~\ref{classes}
the extended version of the data model is closed for union,
intersection and set-difference. \end{property}

\subsection{Normal forms for CSG}

By allowing geometric objects to be constructed from atomic
objects via union, intersection and difference, we arrive at a
situation that is similar to what is used in  the field of
``Constructive Solid Geometry'' (CSG)~\cite{hoffmann}. This is a
method of geometric modeling, where complex static objects are
constructed out of simple objects by taking the union,
intersection and difference.

Looking at literature on CSG, we find that there exists a normal
form for objects composed as Boolean combinations (with the
operators $\cup$, $\cap$, $\setminus$) from atomic objects.

A tree representing a complex object (called a {\em CSG tree}) is
in {\em normal form} when all intersection and subtraction
operators have a left subtree which contains no union operators
and a right subtree which is simply a primitive (a set of polygons
representing a single solid object). All union operators are
pushed towards the root, and all intersection and subtraction
operators are pushed towards the leaves. In our setting, the
primitives are atomic geometric objects and the complexes are
geometric objects.

A CSG tree can be converted to normal form by repeatedly applying
the following set of rewrite rules (which have the Church-Rosser
property) to the tree and then its subtrees:

\begin{center}
\begin{tabular}{llll}
$A \setminus (B \cup C)$ & $ \leadsto $&$(A \setminus B)\setminus
C$&(Rule 1)\\

$A \cap (B \cup C) $&$\leadsto $&$(A \cap B) \cup (A \cap
C)$&(Rule 2)\\

$A \setminus (B \cap C) $&$\leadsto $&$(A \setminus B) \cup (A
\setminus C)$&(Rule 3)\\

$A \cap (B \cap C) $&$\leadsto $&$ (A \cap B) \cap C$&(Rule 4)\\

$A \setminus (B \setminus C) $&$\leadsto$&$ (A \setminus B) \cup
(A \setminus C)$&(Rule 5)\\
 $A \cap (B \setminus C) $&$\leadsto $&$(A \cap B) \setminus
 C$&(Rule 6)\\
 $ (A \setminus B) \cap C $&$\leadsto $&$(A \cap C) \setminus
 B$&(Rule 7)\\
 $(A \cup B) \setminus C $&$\leadsto $&$(A \setminus C) \cup (B \setminus
 C)$&(Rule 8)\\
 $(A \cup B) \cap C $&$\leadsto $&$(A \cap C) \cup (B \cap C)$&(Rule 9)
\end{tabular}

\end{center}

\noindent where $A$, $B$, and $C$ here can be both primitives or
subtrees.

\subsection{Normal forms for geometric objects}

First, we define the notion of normal form for a geometric object
in the extended data model.

\begin{definition}[Normal form]%\label{extendeddatamodel}
\rm
We say that a  {\em geometric object\/} (in the extended version)
is in {\em normal form} if every $\cap$- or $\setminus$-labeled
node has no $\cup$-labeled node in the left subtree and has a
right child that is labeled by an atomic object.\qed
\end{definition}

By Rule 7, differences can be pushed down with respect to
intersections and we obtain, in the set-theoretic notation, that a
geometric object is in normal form if it is of the form
$$\bigcup_{i=1}^n(({\cal O}_{i,1}\cap \cdots \cap {\cal
O}_{i,k_i})\setminus {\cal O}_{i,k_i+1}\setminus \cdots \setminus
{\cal O}_{i,k_i+l_i})$$ where ${\cal O}_{i,j}$ is an atomic
object.

\begin{figure}
\begin{center}
\centerline{\mbox{\psfig{file=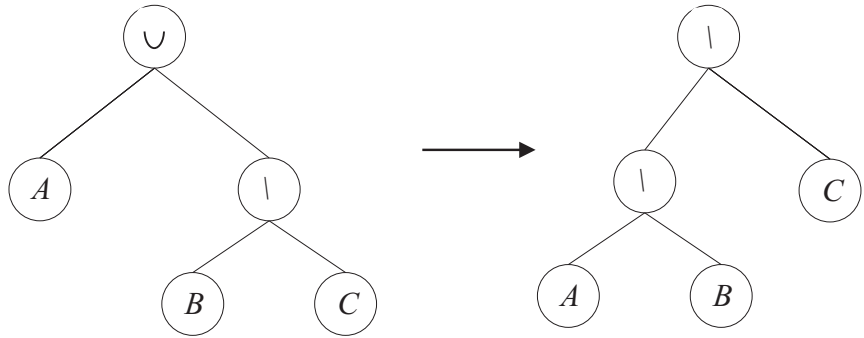}}} \caption[dummy]{Tree
notation for Rule 1. $A$, $B$, and $C$ denote arbitrary subtrees.
The arrow indicates how a subtree can be replaced by another
subtree.} \label{treefig}
\end{center}
\end{figure}

The rewrite Rules 1--9 can be easily converted to tree notation,
as illustrated for Rule 1 in Figure~\ref{treefig}. The following
property says that any geometric object can be rewritten in normal
form. For the proof, we refer to~\cite{gold}.

\begin{property}
Any geometric object in the extended data model can be rewritten,
using Rules 1--9, into a geometric object with the same semantics
that is in normal form. Furthermore, this system of rewrite rules
has the Church-Rosser property.\qed
\end{property}

\section{Conclusion}\label{discussion}

We have introduced the concept of {\it spatio-temporal object} to
model events and objects that change in time. We also specified a
framework for specifying such objects. For some special classes of
{\it spatio-temporal objects} of practical relevance, we
investigated their closure properties with respect to Boolean set
operators. An exhaustive study of these closure properties shows
that the chosen approach leads to mostly negative closure results.
Therefore, we propose an adaptation to the model. The adapted
model has better properties and also is easier to use.

\medskip
To implement our approach, it is sufficient to be able to
represent in a database the following:
\begin{itemize}
\item spatial objects (a solved problem for many classes
of such objects),
\item temporal objects (again a solved problem),
\item function objects (lambda terms).
\end{itemize}
Although to our knowledge none of the currently available DBMS
provides the last option, we believe that the object-relational
(or object-oriented) technology will soon make it feasible. In
fact, one of the earliest object-relational DBMS, Postgres
\cite{Post91}, allowed storing functions as tuple components.
Also, some object-oriented data models, e.g., OODAPLEX
\cite{WuDa93}, permit functions as first-class objects.

Moreover, storing functions themselves is sometimes not necessary.
If the transformation functions are polynomials or rational
functions, they can be represented as lists of coefficients. For
linear polynomials, such lists are of fixed length, opening the
possibility of representing the corresponding spatiotemporal
objects using the standard relational data model.

In addition to implementation issues, it would be challenging to
develop a {\em type system} that captures different {\em
dimensions} of specialization present in geometric objects: region
specialization (polygon, rectangle, ...), transformation
specialization (affine mapping, scaling, ...) and time function
specialization (rational, polynomial, ...).

\bibliographystyle{plain}

\section*{Appendix: Technical proofs from Section 3.2}

\smallskip
\par\noindent
{\bf Proof of Lemma~\ref{polyaffineplus3}.} Let  $(x_1,y_1)$,
$(x_2,y_2)$ and $(x_3,y_3)$ be the three corner points of the
triangle $S$ and let $(x_i,y_i)$ be transformed by  the affinity
$g_i$ given by $$\left(
\begin{array}{@{}c c@{}}a_i(t) & b_i(t)\\ c_i(t) & d_i(t)
\end{array} \right)\left(
\begin{array}{@{}c@{}}x
\\ y \end{array} \right) +  \left( \begin{array}{@{}c@{}} e_i(t)
\\ f_i(t) \end{array} \right), \ \ i=1,2,3.$$

The condition for the existence of a single affine transformation
that transforms these corner points according to their respective
affinities is that the first matrix in the matrix equation below
is regular.

$$ \left(
\begin{array}{@{}cccccc@{}}
x_1 & y_1 & 0 & 0 & 1 &0\\ 0 &0 & x_1 & y_1 & 0 & 1\\ x_2 & y_2 &
0 & 0 & 1 &0\\ 0 &0 & x_2 & y_2 & 0 & 1\\ x_3 & y_3 & 0 & 0 & 1
&0\\ 0 &0 & x_3 & y_3 & 0 & 1\\
 \end{array}
 \right) \left(
\begin{array}{@{}c@{}}
a(t)\\  b(t) \\ c(t)\\  d(t)\\ e(t) \\ f(t) \\
 \end{array}
 \right) = \left(
\begin{array}{@{}c@{}}
a_1(t)x_1+b_1(t)y_1+e_1(t)\\  c_1(t)x_1+d_1(t)y_1+f_1(t) \\
a_2(t)x_2+b_2(t)y_2+e_2(t)\\  c_2(t)x_2+d_2(t)y_2+f_2(t)\\
a_3(t)x_3+b_3(t)y_3+e_3(t)
\\ c_3(t)x_3+d_3(t)y_3+f_3(t)
\\
 \end{array}
 \right)$$

This is the case if and only if the three points $(x_1,y_1)$,
$(x_2,y_2)$ and $(x_3,y_3)$ are not collinear. By assumption, this
condition is satisfied. We find the affine transformation that
transforms the triangle $S$ according to the different movements
of the corner points, by solving the above matrix equation.

The result of this computation is the affine transformation with
coefficients $a(t)$, $b(t)$, $c(t)$, $d(t)$, $e(t)$, and $f(t)$
that have the following form (to save space time dependence is
omitted): { $$\begin{array}{lll} a(t)&=&
\frac{-x_1a_1y_3+x_1a_1y_2-y_2b_3y_3+e_2y_3-e_1y_3-y_2e_3+e_1y_2+y_1a_3x_3}
{x_1y_2-x_1y_3+x_2y_3-x_3y_2+x_3y_1-x_2y_1}\\ &&
  +
\frac{a_2x_2y_3+y_1e_3-y_2a_3x_3+b_2y_2y_3-y_1a_2x_2+y_1b_3y_3-y_1e_2-y_1b_2y_2+b_1y_1y_2-b_1y_1y_3}
{x_1y_2-x_1y_3+x_2y_3-x_3y_2+x_3y_1-x_2y_1},\\ &&
\\
b(t)&=&
-\frac{-x_1b_2y_2-x_2a_3x_3+x_2b_1y_1+x_1b_3y_3-x_3a_1x_1-x_1a_2x_2-x_2b_3y_3-x_2e_3+x_3a_2x_2}
{x_1y_2-x_1y_3+x_2y_3-x_3y_2+x_3y_1-x_2y_1}\\ &&

-
\frac{-x_3e_1+x_3b_2y_2+x_1a_3x_3+x_2a_1x_1+x_1e_3+x_2e_1-x_3b_1y_1+x_3e_2-x_1e_2}
{x_1y_2-x_1y_3+x_2y_3-x_3y_2+x_3y_1-x_2y_1},\\ &&
\\
c(t)&=&
\frac{x_1c_1y_2-x_1c_1y_3-y_1c_2x_2+y_1f_3-y_2f_3+f_1y_2+y_1c_3x_3-y_2d_3y_3+y_1d_3y_3}{x_1y_2-x_1y_3+x_2y_3-x_3y_2+x_3y_1-x_2y_1}
\\ &&
   +
\frac{-y_1d_2y_2+d_1y_1y_2+f_2y_3-y_2c_3x_3-y_1f_2-d_1y_1y_3+c_2x_2y_3+d_2y_2y_3-f_1y_3}{x_1y_2-x_1y_3+x_2y_3-x_3y_2+x_3y_1-x_2y_1},\\
\\
 d(t) &=&  -\frac{x_2d_1y_1+x_1d_3y_3-x_3d_1y_1+x_3d_2y_2-x_1c_2x_2+x_3c_2x_2+x_2c_1x_1-x_2f_3-x_3f_1}{x_1y_2-x_1y_3+x_2y_3-x_3y_2+x_3y_1-x_2y_1}\\
 && -
\frac{x_1c_3x_3-x_2c_3x_3-x_2d_3y_3-x_1d_2y_2-x_3c_1x_1+x_2f_1+x_1f_3-x_1f_2+x_3f_2}{x_1y_2-x_1y_3+x_2y_3-x_3y_2+x_3y_1-x_2y_1},\\
\\
e(t) &=&
\frac{y_1x_3e_2+y_2x_1e_3-y_2x_3e_1-y_2x_3a_1x_1+y_2x_1b_3y_3-e_2x_1y_3+e_1x_2y_3+b_1y_1x_2y_3-a_2x_2x_1y_3}{x_1y_2-x_1y_3+x_2y_3-x_3y_2+x_3y_1-x_2y_1}\\
&& +
\frac{-y_1x_2e_3-b_2y_2x_1y_3+y_2x_1a_3x_3+y_1x_3b_2y_2+y_1x_3a_2x_2-y_1x_2b_3y_3-y_1x_2a_3x_3-y_2x_3b_1y_1+a_1x_1x_2y_3}{x_1y_2-x_1y_3+x_2y_3-x_3y_2+x_3y_1-x_2y_1},\\
\\
f(t)&=&
\frac{-y_2x_3f_1-y_2x_3c_1x_1+y_1x_3d_2y_2+y_1x_3c_2x_2-y_1x_2c_3x_3-y_1x_2d_3y_3+c_1x_1x_2y_3+y_2x_1f_3-y_1x_2f_3}{x_1y_2-x_1y_3+x_2y_3-x_3y_2+x_3y_1-x_2y_1}\\
 && +
\frac{y_1x_3f_2+f_1x_2y_3-f_2x_1y_3+y_2x_1c_3x_3-y_2x_3d_1y_1+y_2x_1d_3y_3+d_1y_1x_2y_3-d_2y_2x_1y_3-c_2x_2x_1y_3}{x_1y_2-x_1y_3+x_2y_3-x_3y_2+x_3y_1-x_2y_1}.\\
\end{array}$$
}

Indeed, the transformation matrix
$$\left(\begin{array}{@{}cc@{}}a(t) & b(t)\\ c(t) & d(t)\\
\end{array}\right) $$ is regular. Simplifying the expression $a(t)d(t)-b(t)c(t)$
gives the result $$\frac{x_1(t)y_2(t) - x_2(t)y_1(t) -
x_1(t)y_3(t) + x_3(t)y_1(t) + x_2(t)y_3(t) - x_3(t)y_2(t)}{y_2x_1
- y_3x_1 - y_2x_3 + y_1x_3 + y_3x_2 - y_1x_2},$$ where $x_i(t) =
a_i(t)x_i + b_i(t)y_i + e_i(t)$ and $y_i(t)
=
c_i(t)x_i + d_i(t)y_i + f_i (t)$, $i = 1,2,3.$ This denominator of
this expression is zero if and only if the three points
$(x_1,y_1)$, $(x_2,y_2)$ and $(x_3,y_3)$ are collinear. By
assumption, the points $(x_1,y_1)$, $(x_2,y_2)$ and $(x_3,y_3)$
form a triangle, however. The numerator is non-zero since the
points $g_1(x_1,y_1;t)$, $g_2(x_2,y_2;t)$ and $g_3(x_3,y_3;t)$
form a triangle $S_t$ at any moment $t\in I$.

The coefficients of the resulting affine transformation are linear
functions of the coefficients of the original transformations of
the corner points $(x_1,y_1)$, $(x_2,y_2)$ and $(x_3,y_3)$. As the
original transformations are rational, the resulting affine
transformation is rational too. \qed

\medskip

\smallskip
\par\noindent
{\bf Proof of Lemma~\ref{polyaffineplus2}.} Let  $(x_1,y_1)$ and
$(x_2,y_2)$ be the two end points of the line segment  $L$ and let
$(x_i,y_i)$ be transformed by the affinity $g_i$ given by $$\left(
\begin{array}{@{}c c@{}}a_i(t) & b_i(t)\\ c_i(t) & d_i(t)
\end{array} \right)\left(
\begin{array}{@{}c@{}}x
\\ y \end{array} \right) +  \left( \begin{array}{@{}c@{}} e_i(t)
\\ f_i(t) \end{array} \right), \ \ i=1,2.$$

 We
prove that there always exists a rational affine functions $a(t)$,
$b(t)$, $c(t)$ and $d(t)$, such that the matrix
$$\left(\begin{array}{@{}cc@{}}a(t) & b(t)\\ c(t) & d(t)\\
\end{array}\right) $$
transforms the line segment as described in the statement of this
lemma (so, the translation components $e(t)$ and $f(t)$ of this
affinity are identical zero).

The condition for the existence of a single affinity that
transforms the two endpoints of the line segment according to
their respective affinities is that the first matrix in the
following equation is regular.

$$\left(\begin{array}{@{}cccc@{}}x_1 & y_1 & 0 & 0 \\ 0 & 0 & x_1
& y_1
\\ x_2 & y_2 & 0 & 0  \\0 & 0 & x_2 & y_2 \\
\end{array}\right)  \left(\begin{array}{@{}c@{}}a(t) \\ b(t)\\ c(t)  \\d(t) \\
\end{array}\right)= \left(\begin{array}{@{}c@{}}a_1(t)x_1+b_1(t)y_1+e_1(t)
\\c_1(t)x_1+d_1(t)y_1+f_1(t) \\ a_2(t)x_2+b_2(t)y_2+e_2(t) \\ c_2(t)x_2+d_2(t)y_2+f_2(t)\\
\end{array}\right) $$

This is true if the two endpoints of the line segment do not
coincide.

The affinity that determines the movement of the intersection, can
be found by solving the above equation: it is given by  $$
\begin{array}{l}a(t)
=
\frac{e_1(t)y_2-y_1a_2(t)x_2-y_1b_2(t)y_2-y_1e_2(t)+a_1(t)x_1y_2+b_1(t)y_1y_2}{x_1y_2-x_2y_1}
\\
\\b(t) = -\frac{x_2e_1(t)-x_1a_2(t)x_2-x_1b_2(t)y_2-x_1e_2(t)+x_2a_1(t)x_1+x_2b_1(t)y_1}{x_1y_2-x_2y_1}
\\
 \\ c(t) =  \frac{f_1(t)y_2-y_1c_2(t)x_2-y_1d_2(t)y_2-y_1f_2(t)+c_1(t)x_1y_2+d_1(t)y_1y_2}{x_1y_2-x_2y_1}
\\
 \\ d(t) = -\frac{x_2f_1(t)-x_1c_2(t)x_2-x_1d_2(t)y_2-x_1f_2(t)+x_2c_1(t)x_1+x_2d_1(t)y_1}{x_1y_2-x_2y_1}. \\
\end{array}$$

 As in the
case of the previous lemma, it can be shown that
$$\left(\begin{array}{@{}cc@{}}a(t) & b(t)\\ c(t) & d(t)\\
\end{array}\right) $$ is regular and therefore determines an affinity.

This solution is linear in the components of the original rational
affine transformations of ${\cal O}_1$ and ${\cal O}_1$, so it is
also rational. \qed

\medskip

\par\noindent
{\bf Proof of Lemma~\ref{intersection2lines}.} Let  $(x_i,y_i)$
and $(u_i,v_i)$ be the two end points of the line segment  $L_i$
($i=1,2$). Let $L_i$ be transformed by the affinity $g_i$ given by
$$\left(
\begin{array}{@{}c c@{}}a_i(t) & b_i(t)\\ c_i(t) & d_i(t)
\end{array} \right)\left(
\begin{array}{@{}c@{}}x
\\ y \end{array} \right) +  \left( \begin{array}{@{}c@{}} e_i(t)
\\ f_i(t) \end{array} \right), \ \ i=1,2.$$

\noindent We compute the intersection of $g_1(L_1;t)$ and
$g_2(L_2;t)$ by solving the equations

\smallskip

\begin{tabular}{l}
$\lambda_1 (a_1(t)x_1 + b_1(t)y_1 + e_1(t)) + (1 - \lambda_1)
(a_1(t)u_1 + b_1(t)v_1 + e_1(t)) =$ \\ $ \lambda_2 (a_2(t)x_2 +
b_2(t)y_2 + e_2(t)) + (1 - \lambda_2) (a_2(t)u_2 + b_2(t)v_2 +
e_2(t))$ \\ and \\ $\lambda_1 (c_1(t)x_1 + d_1(t)y_1 + f_1(t)) +
(1 - \lambda_1) (c_1(t)u_1 + d_1(t)v_1 + f_1(t))=$ \\ $\lambda_2
(c_2(t)x_2 + d_2(t)y_2 + f_2(t)) + (1 - \lambda_2) (c_2(t)u_2 +
d_2(t)v_2 + f_2(t))$
\end{tabular}

in $\lambda_1$ and $\lambda_2$. The determinant of the matrix
$$\left( \begin{array}{@{}c c@{}}a_1(t)x_1 + b_1(t)y_1 -a_1(t)u_1
- b_1(t)v_1 & a_2(t)u_2 + b_2(t)v_2 -a_2(t)x_2 - b_2(t)y_2\\
c_1(t)x_1 + d_1(t)y_1 -c_1(t)u_1 - d_1(t)v_1 & c_2(t)u_2 + d_2(t)
- v_2 c_2(t)x_2 - d_2(t)y_2\end{array} \right)$$ is zero if one of
the $g_i(L_i;t)$ is parallel to one of the coordinate axes or if
both line segments are parallel. The latter case is no problem as
we can use the finite time partition
(Property~\ref{finitetimepartition}) to consider only those
subintervals $J$ of  $I$ during which the intersection exists. We
treat the case of line segments parallel to one of the coordinate
axis separately.

If the line segments are not parallel to one of the coordinate
axes, the intersection point is the following. We only give the
$x$-coordinate $s_x(t)$ (the $y$-coordinate $s_y(t)$ is expressed
similarly). For clarity time dependence in the coefficients of the
affinities is omitted.

We have that $s_x(t)((a_1x_1 + b_1y_1 -a_1u_1 - b_1v_1)(-d_2v_2+
c_2x_2+d_2y_2-c_2u_2) +(a_2u_2 + b_2v_2 -a_2x_2
-b_2y_2)(c_1x_1+d_1y_1-c_1u_1-d_1v_1)) $ equals

\medskip

{\small \hspace{-1cm}
\begin{tabular}{l}
$((((x_2v_1-u_2v_1)a_2+(-v_2v_1+y_2v_1)b_2)d_1+((f_1-d_2v_2-f_2)x_2+u_2d_2y_2+(-f_1+f_2)u_2)a_2+$\\

$(e_2c_2+b_2v_2c_2)x_2+((-f_2-c_2u_2+f_1)y_2+(-f_1+f_2)v_2)b_2+e_2d_2y_2-e_2d_2v_2-e_2c_2u_2)x_1+$\\

$((u_2u_1-x_2u_1)a_2+(-y_2u_1+v_2u_1)b_2)d_1y_1+((u_1d_2v_2+(-f_1+f_2)u_1)x_2-u_2u_1d_2y_2+$\\

$(f_1-f_2)u_2u_1)a_2+(-b_2v_2u_1c_2-e_2u_1c_2)x_2+((u_1c_2u_2+(-f_1+f_2)u_1)y_2+(f_1-f_2)v_2u_1)b_2-$\\

$e_2u_1d_2y_2+e_2u_1d_2v_2+e_2u_1c_2u_2)a_1+(((-x_2v_1+u_2v_1)a_2+(-y_2v_1+v_2v_1)b_2)c_1b_1+$\\
$((u_2e_1-x_2e_1)a_2+(v_2e_1-y_2e_1)b_2)c_1)x_1+((((-u_2u_1+x_2u_1)a_2+(-v_2u_1+y_2u_1)b_2)c_1+$\\
$((f_1-d_2v_2-f_2)x_2+u_2d_2y_2+(-f_1+f_2)u_2)a_2+(e_2c_2+b_2v_2c_2)x_2+$\\
$((-f_2-c_2u_2+f_1)y_2+(-f_1+f_2)v_2)b_2+e_2d_2y_2-e_2d_2v_2-e_2c_2u_2)y_1+$\\
$((v_1d_2v_2+(-f_1+f_2)v_1)x_2-u_2v_1d_2y_2+(f_1-f_2)u_2v_1)a_2+(-b_2v_2v_1c_2-e_2v_1c_2)x_2+$\\
$((v_1c_2u_2+(-f_1+f_2)v_1)y_2+(f_1-f_2)v_2v_1)b_2-e_2v_1d_2y_2+e_2v_1c_2u_2+e_2v_1d_2v2)b_1+$\\
$((u_2e_1-x_2e_1)a_2+(v_2e_1-y_2e_1)b_2)d_1y_1+((x_2u_1e_1-u_2u_1e_1)a_2+$\\
$(y_2u_1e_1-v_2u_1e_1)b_2)c_1+((x_2v_1e_1-u_2v_1e_1)a_2+(-v_2v_1e_1+y_2v_1e_1)b_2)d_1)$.
\end{tabular}}

\medskip

For the intersection point to exists, $((a_1x_1 + b_1y_1 -a_1u_1 -
b1_v1)(-d_2v_2+ c_2x_2+d_2y_2-c_2u_2) +(a_2u_2 + b_2v_2 -a_2x_2
-b_2y_2)(c_1x_1+d_1y_1-c_1u_1-d_1v_1)) $ should be different from
zero. This condition expresses the fact that the line segments are
not parallel, which is true by assumption.

The intersection point moves rationally, as its functions of time
are rational functions in the coefficients of the original
transformations. For any choice of reference point, it is clear
that a rational affinity can be found that moves it as described
by the above formulas $(s_x(t),s_y(t))$.

If one of the line segments $g_1(L_1;t)$ or $g_2(L_2;t)$ is
parallel to the $x$-axis, the intersection point will have as
$y$-coordinate the $y$-coordinate of that line segment. The same
holds for segments parallel to the $y$-axis. In the case that one
segment is parallel to the $y$-axis and the other to the $x$-axis,
 the intersection point moves with linear, polynomial, respectively rational functions of time,
if both the objects ${\cal O}_1$ and ${\cal O}_2$ move with
linear, polynomial, respectively rational functions of time.
 \qed

\end{document}